\documentclass[epj]{svjour}
\usepackage{amsmath,amssymb,epsfig,bm,mathrsfs,slashed,color,graphicx,mathtools,isotope,aas_macros}%,feynmp}
\usepackage{hyperref}
\hypersetup{colorlinks=true, linkcolor=red, filecolor=magenta, urlcolor=red, citecolor=blue}
\usepackage[numbers,sort&compress]{natbib}
\usepackage{filecontents}
\usepackage{tikz}
\usepackage[utf8]{inputenc}
\definecolor{red}{rgb}{0.8,0,0}
\definecolor{violet}{rgb}{0.4,0,0.4}
\definecolor{green}{rgb}{0,0.5,0.0}
\definecolor{navy}{rgb}{0.0,0.0,0.6}
\definecolor{orange}{rgb}{0.8,0.2,0.0}
\newcommand{\bea}{\begin{eqnarray}}
\newcommand{\eea}{\end{eqnarray}}
\newcommand{\ep}{\epsilon}

\newcommand{\vecp}{{\bm p}}
\newcommand{\vecq}{{\bm q}}
\newcommand{\veck}{{\bm k}}
\newcommand{\vecv}{{\bm v}}
\newcommand{\vecE}{{\bm E}}

\newcommand{\vecj}{{\bm j}}

\newcommand{\vecB}{{\bm B}}

\newcommand{\ie}{{\it i.e.}}

\journalname{Eur. Phys. J. A}

\begin{document}
\title{
Thermal conduction and thermopower of a warm neutron star crust in magnetic fields
}

\author{Henrik Danielyan$^1$,  Arus Harutyunyan$^{1,2}$, Armen Sedrakian$^{3,4}$ }
\institute{
$^1$   Physics Institute, Yerevan State University, Yerevan 0025, Armenia \\
$^2$ Byurakan Astrophysical Observatory,  Byurakan 0213, Armenia\\
$^3$ Institute of Theoretical Physics, University of Wroc\l{}aw, pl. M. Borna 9,   50-204 Wroc\l{}aw, Poland\\
$^4$  Frankfurt Institute for Advanced Studies, D-60438 Frankfurt am Main, Germany\\
}
\date{Received: 9 May 2026  / Accepted: 25 August 2026\\
\textcopyright \, 
The Author(s),  under exclusive licence to Società Italiana di Fisica and Springer-Verlag GmbH Germany, part of Springer Nature  2026
\\
Communicated by S. Typel 
}

\abstract{
  We compute the thermal conductivity and thermoelectric power
  (thermopower) of the inner crust of compact stars across a broad
  temperature–density domain relevant for proto–neutron stars, binary
  neutron-star mergers, and accreting neutron stars. The analysis
  covers the transition from a semi-degenerate to a highly degenerate
  electron gas and assumes temperatures above the melting threshold of
  the nuclear lattice, such that nuclei form a liquid. The nuclear
  composition is prescribed by selected zero-temperature inner-crust
  models and is kept fixed when the temperature is varied. The transport
  coefficients are obtained by solving the Boltzmann kinetic equation
  in the relaxation-time approximation, fully incorporating the
  anisotropies generated by non-quantizing magnetic fields.  Electron
  scattering rates include (i) dynamical screening of the electron–ion
  interaction in the hard-thermal-loop approximation of QED, (ii)
  ion–ion correlations within a one-component plasma, and (iii) finite
  nuclear-size effects. As an additional refinement, we evaluate
  electron–neutron scattering induced by the coupling of electrons to
  the anomalous magnetic moment of free neutrons; this contribution is
  found to be subdominant throughout the parameter range explored.  To
  assess the sensitivity of transport coefficients to the underlying
  microphysics, we perform calculations for several inner-crust
  compositions obtained from different nuclear interactions and
  many-body methods. Across most of the crust, variations in
  relaxation times and in the components of the anisotropic
  thermal-conductivity and thermopower tensors reach up to factors
  3 to 4 and 1.5 to 2, respectively, with the exception of the
region where pasta phases are expected. Within the adopted
  single-component, fixed-composition approximation, these results
  provide a reference set of composition-dependent microphysical inputs
  for dissipative magneto-hydrodynamic simulations of warm neutron stars
  and post-merger remnants, where anisotropic heat and charge transport
  are of critical importance.
}
\PACS{
      {26.60.Gj}{Neutron star crust} \and
      {52.25.Fi}{Transport properties} \and
      {52.25.Xz}{Magnetized plasmas} \and
      {97.60.Jd}{Neutron stars}
     }

\authorrunning {H. Danielyan et al.}
\titlerunning{ Thermal conduction and thermopower of a neutron star crust }
\maketitle

\section{Introduction}
\label{sec:intro}

Transport coefficients of cold and dense plasma in neutron stars, both
in the liquid and solid phases, have been extensively studied for many
decades
~\cite{Flowers1976,Yakovlev1980,Urpin1980,Flowers1981,Itoh1983,Itoh1984,Nandkumar1984MNRAS,Sedrakian1987,Itoh1993,Baiko1998,Potekhin1999,Shternin2006,Itoh2008}
(see Refs.~\cite{Schmitt2018, Potekhin2015} for reviews). More
recently, attention has shifted toward transport in the
high-temperature regime, where a transition from cold, degenerate
matter to semi-degenerate or non-degenerate plasma may occur. Dense
nuclear matter in such heated plasma states is primarily realized in
binary neutron star mergers and in proto-neutron stars formed during
supernova explosions. In these environments, matter is expected to
reach temperatures of order 100~MeV, accompanied by strong thermal and
compositional gradients and intense electromagnetic fields. Neutron
stars may also be heated through accretion; however, the temperatures
achieved in this case are typically one to two orders of magnitude
lower than those reached in these transient events~\cite{Jain2025}.

The primary focus of our work is the transport properties of the {\it
  inner crust} of neutron stars, where ultrarelativistic electrons,
heavy nuclei, and a Fermi gas of unbound neutrons coexist. In this
region, electrons dominate both charge and heat transport, as the
contribution of normal (non-superfluid) neutrons to the thermal
conductivity is known to be negligible~\cite{Bisnovatyi-Kogan1982}. A
detailed calculation of the electrical conductivity in this
inner-crust regime was recently carried out in
Ref.~\cite{Harutyunyan2024-1}.  Complementary studies of {\it
  outer-crust} matter -- including the electrical and thermal
conductivities of hot, dense, fully ionized plasmas -- were performed
in Refs.~\cite{Harutyunyan2016,Harutyunyan2024-2,Petrosyan2026}. 
The electrical conductivity results of Ref.~\cite{Harutyunyan2016} 
were subsequently used to quantify the roles of Ohmic dissipation 
and the Hall effect in binary neutron-star mergers~\cite{Harutyunyan2018}, 
and to delineate the parameter space in which a magnetohydrodynamic 
treatment remains valid. These outer-crust studies provide a useful 
reference, while the more recent extension to the inner crust~\cite{Harutyunyan2024-1} addresses the additional microphysics 
introduced by the presence of free neutrons and heavier nuclear clusters.

The dominant scattering channel, which determines the electron
transport in the inner crust, is the Coulomb scattering off the
ions. However, electron-neutron interaction via the coupling of the
electron to the neutron's anomalous magnetic moment has also been
studied. Refs.~\cite{Flowers1976, Flowers:1979} considered the
low-temperature regime where the neutron star matter is in a solid
state in the crust or forms a Fermi-liquid in the core. It was found
that the effect of electron-neutron scattering on the thermal
conductivity is small in the inner crust~\cite{Flowers1976}, but can
become important in the core~\cite{Flowers:1979}.
Ref.~\cite{Bertoni2015} studied the induced electron-neutron
interaction due to ions (in the inner crust) or protons (in the core),
and found that the contribution of this interaction to transport
coefficients is always negligible in the inner crust.

In the presence of magnetic fields, the conduction in the crust
becomes anisotropic, and the transport in the direction transverse to
the magnetic field becomes suppressed. Additionally, the simultaneous
presence of electromagnetic fields and thermal gradients mixes various
transport channels, leading to coupled non-trivial dynamics in the
evolution of temperature and magnetic field profiles.  In particular,
a generation mechanism of strong magnetic fields via the dynamo effect
(\ie, magnetic field amplification) driven by strong thermal gradients
was proposed earlier in
Refs.~\cite{Dolginov1980,Blandford1983,Geppert1991,Wiebicke1996,Gakis2024}.
Thermoelectric and thermomagnetic effects have also been considered in
the context of thermal evolution of neutron
stars~\cite{Kaminker2006,Geppert2017,Dehman2023,Ascenzi2024}.

In this work, we will compute the thermal conductivity and the
thermoelectric power (thermopower) in the inner crust of a moderately
warm neutron star, including the anisotropies of the transport induced
by strong magnetic fields below the quantization limit $B\le
10^{14}$~G. We will extend the formalism developed previously for
assessment of electrical and thermal conductivities in the outer
crust~\cite{Harutyunyan2016,Harutyunyan2024-2} to the inner crust,
adopting five composition models for spherical
nuclei~\cite{Negele1973,Mondal2020,Pearson:2018,Raduta2019}, which
were employed recently to compute the electrical conductivity of the
same system in Ref.~\cite{Harutyunyan2024-1}. 

The range of densities considered in this work is bounded from below
by the neutron drip density
$\log_{10}(\rho_{\rm d}/\mathrm{g\,cm^{-3}}) \simeq 11.6$ and from
above by the onset density of nuclear pasta phases $\rho_p$, estimated
to lie in the range
$\log_{10}(\rho_{p}/\mathrm{g\,cm^{-3}}) \simeq
13.5–13.9$~\cite{Oyamatsu1993,CaplanHorowitz2017}.  Simulations
indicate that pasta-like density inhomogeneities may survive up to
$T\gtrsim 1$~MeV -- the lower range of temperatures considered here,
but thermal fluctuations wash out long-range order, making a
well-defined lattice
questionable~\cite{Avancini2010,Caplan2015,Schneider2018}. These
disordered structures can still survive and affect
transport~\cite{Sonoda2007,HorowitzBerry2008,Horowitz2015,Pelicer2023}.

The scope of the microscopic treatment should be kept in mind in
interpreting the results. First, ion--ion correlations are described by
the \emph{static} structure factor $S(q)$ of a classical
one-component Coulomb plasma (OCP). Thus the ions are treated as a
single species for their correlations and energy transfer associated
with ionic motion is neglected in the electron--ion collision kernel.
A treatment of dynamical correlations would instead require the
dynamical structure factor $S(q,\omega)$. Second, finite nuclear size
is included only in the electron--ion vertex through the single-ion
charge form factor $F(q)$, whereas the OCP structure factor itself is
that of point ions. Consequently, modifications of the ion--ion
correlations and of the melting criterion caused by finite charge
extent or screened ion--ion interactions are not included. Finally,
the five adopted compositions are single-species, zero-temperature
compositions. Finite-temperature multi-component calculations show
that a distribution of nuclear species can develop near
crystallization and in a hot proto-neutron-star crust
~\cite{Carreau2020MCP,DinhThi2023MCP}. Accordingly, the calculations
at the upper end of our temperature interval, in particular
$T\simeq10$~MeV, should be regarded as a {\it fixed-composition
extrapolation} of the transport formalism rather than as a
self-consistent prediction of the equilibrium crust composition.
These approximations are expected to be most restrictive in the
deepest inner crust, close to the pasta region, and in matter with a
broad nuclear distribution such as may occur in accreting or merger
environments.

In addition to electron–ion scattering, we also include the full
finite-temperature scattering rate of electrons off neutrons via the
neutron’s anomalous magnetic moment. Although earlier studies
concluded that electron–neutron scattering is
negligible~\cite{Flowers1976}, those calculations were performed
assuming exchange of longitudinally screened plasmons with Debye
screening. As we show below, the dominant contribution instead arises
from the exchange of transversely screened plasmons. The longitudinal
and transverse components of the photon polarization tensor are
evaluated within the hard-thermal-loop (HTL) effective field
theory~\cite{Braaten1990,Braaten1992,Harutyunyan2016}.  Finally, we
{\it provide order-of-mag\-ni\-tude estimates for the evolution timescales}
of crustal magnetic fields and identify the conditions under which
thermoelectric effects may play a significant role in their dynamics.

This work is organized as follows. Section~\ref{sec:regimes} discusses
briefly the physical conditions in the inner crust of a neutron star
for five different compositions employed in this
work. Section~\ref{sec:Boltzmann} provides the derivation of the
tensors of electrical and thermal conductivities and thermopower in
magnetic fields from the Boltzmann equation. Section~\ref{sec:results}
presents the numerical results for the thermal conductivity,
thermopower, and magnetic field evolution timescales in the density,
temperature, and $B$-field regimes of interest. Our results are
summarized in Section~\ref{sec:summary}.  Appendix~\ref{app:matrix_en}
derives the electron-neutron scattering matrix element in a thermal
medium, and Appendix~\ref{app:relax_en} computes the resulting
electron relaxation time.  Appendix~\ref{app:low-T} collects the
low-temperature formulae for the thermoelectric coefficients. We use
the natural units with $\hbar = c = k_B = k_e = 1$, and the metric
signature $(1,-1,-1,-1)$.

\section{Physical conditions in inner crust}
\label{sec:regimes}

The inner crust of a neutron star consists of relativistic electrons,
fully ionized nuclei, and a gas of unbound neutrons that emerges once
the density exceeds the neutron-drip threshold. The total baryon
density in the inner crust is then given by
%-----------------------------------
\bea\label{eq:n_B}
n_B = An_i + (1-V_Nn_i)n_n,
\eea
%-----------------------------------
where $n_i$ and $n_n$ are the number densities of ions and unbound
neutrons, respectively, $V_N$ is the volume of the nucleus, and
$V_Nn_i$ is the excluded volume correction~\cite{Baym1971}. The state
of ions in the crust is characterized by the plasma parameter
%-----------------------------------
\bea\label{eq:Gamma}
\Gamma = \frac{e^2 Z^2}{TR_{\rm WS}},
\eea
%-----------------------------------
where $e$ is the elementary charge, $Z$ is the ion charge number, $T$
is the temperature, $R_{\rm WS}=(4\pi n_i/3)^{-1/3}$ is the radius of
the spherical volume per ion, \ie, that of the Wigner-Seitz cell.
Within the classical point-ion OCP model, ions form a weakly coupled
Boltzmann gas for $\Gamma\ll 1$, a strongly coupled liquid for
$1\le\Gamma\le \Gamma_m\simeq 160$, and a solid for
$\Gamma>\Gamma_m$. The corresponding OCP melting temperature is
$T_m=(e^2 Z^2)/(\Gamma_m R_{\rm WS})$. Screening, finite charge extent,
and a multi-component composition can modify the phase boundary, so
$\Gamma_m\simeq160$ is used here as the reference criterion defining
the liquid regime rather than as a model-independent crystallization
condition. The current transport calculation applies to this liquid
regime.

For numerical computations, we will adopt five different
density-dependent compositions of crustal matter, which we label as
NV~\cite{Negele1973}, D1M and D1M*~\cite{Mondal2020},
Bsk24~\cite{Pearson:2018}, and Sly9~\cite{Raduta2019}. These
compositions were computed at $T=0$ and, as a deliberate
simplification, we keep their $Z$, $A$, nuclear radii, and
free-neutron fractions fixed when varying the temperature. This
isolates the temperature dependence of the electron transport problem
for a prescribed ionic background, but it does not include thermal
rearrangement of the nuclear composition. Finite-temperature
multi-component calculations near
crystallization~\cite{Carreau2020MCP} and for proto-neutron-star inner
crusts~\cite{DinhThi2023MCP} show that thermal population of multiple
nuclear species can be important. Thus the $T=10$~MeV results should
be read as an illustrative fixed-composition extrapolation; a
self-consistent treatment at such temperatures requires a
finite-tem\-pera\-ture composition and, potentially, cluster
dissolution, which is outside the present scope. The key properties of
these compositions, such as the nuclear charge $Z$, mass number $A$,
and the fraction of unbound neutrons $Y_n = n_n/n$, where $n_n$ and
$n$ are the neutron and total number densities, were analyzed and
compared extensively in Ref.~\cite{Harutyunyan2024-1}. For
completeness, we summarize the main features here.

The primary distinctions among the five compositions stem from
differences in the predicted mass number $A$ at high densities,
specifically in the range $13.5 \le \log_{10}\rho\,[\mathrm{g\,cm^{-3}}]
\le 14$, where $\rho$ denotes the matter mass density. The D1M and
D1M* models yield comparatively large nuclei, with $300 \le A \le
2000$, accompanied by a low free–neutron fraction $0.7 \ge Y_n \ge
0.2$ in this density interval. In contrast, the Bsk24 and Sly9 models
predict much smaller nuclei with $A \sim 200$ and a correspondingly
higher neutron fraction $Y_n \sim 0.8$. The NV composition shows an
intermediate behavior between these two classes for $\log_{10}\rho$ [g cm$^{-3}] \ge 13.5$.
%-------------------------------------------------------
\begin{figure}[thb] 
\begin{center}
\includegraphics[width=\linewidth,keepaspectratio]{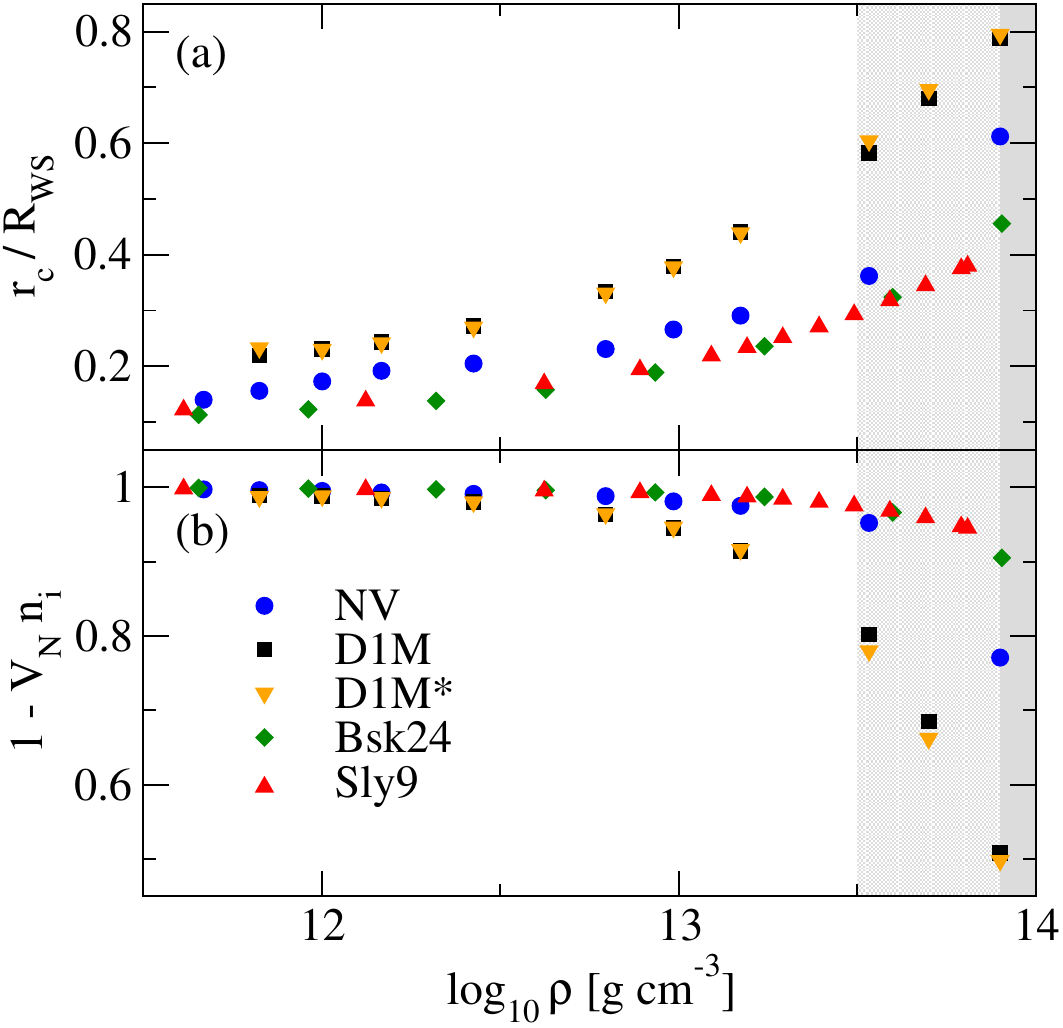}
\caption{(a)~Ratio of the nuclear radius $r_c$ to the Wigner–Seitz
  cell radius and (b) the volume fraction $1 - V_N n_i$ occupied by
  the neutron gas, shown as functions of density for five compositions
  of stellar matter from
  Refs.~\cite{Negele1973,Mondal2020,Pearson:2018,Raduta2019}.  The
  light shaded region indicates the density range
  $\log_{10}(\rho_p/\mathrm{g\,cm^{-3}}) \simeq 13.5-13.9$, where our
  approach breaks down due to either the onset of finite-temperature
  pasta phases, as expected according to various
  calculations~\cite{CaplanHorowitz2017}, or the breakdown of the
  expansion in terms of small parameter $r_c / R_{\mathrm{WS}}$. The
  dark shaded region with high certainty corresponds to the pasta
  phase.}
\label{fig:radii} 
\end{center}
\end{figure}
%-------------------------------------------------------

Figure~\ref{fig:radii} shows the ratio of the radius $r_c$ of the
nucleus in the ground state to that of the Wigner-Seitz cell, as well
as the free-volume fraction for neutrons, $(1-V_Nn_i)$, as a function
of density. At high densities, the D1M and D1M* models predict
extremely large nuclei that occupy a substantial fraction of the cell
volume. Consequently, these models leave significantly less available
volume for the free–neutron component compared with the other
compositions.  The light-shaded region in Fig.~\ref{fig:radii} marks
the density range where our approach is expected to lose validity for
two reasons. First, the possible onset of nuclear pasta
phases~\cite{Oyamatsu1993,CaplanHorowitz2017} invalidates the
assumption of spherical nuclei; changes in nuclear shape and topology
may lead to qualitatively different transport properties of pasta
matter. Second, the transport treatment becomes unreliable when the
parameter $r_c/R_{\mathrm{WS}}$, where $r_c$ is the nuclear charge
radius and $R_{\mathrm{WS}}$ is the Wigner–Seitz cell radius, is no
longer small. This latter condition is encountered for the D1M and
D1M* models, but not for the remaining models considered here. At
leading order in $r_c/R_{\mathrm{WS}}$, nuclei may be treated as
effective point ions for the purpose of ion–ion correlations, while
finite nuclear size effects are incorporated solely through the
single-ion charge form factor $F(q)$.

In the regime of interest, electrons form a degenerate or
semi-degenerate ultrarelativistic Fermi gas, with the Fermi energy
given by $\ep_F= (p_F^2+m_e^2)^{1/2}\geq 25$~MeV, $p_F =
(3\pi^2n_e)^{1/3}$, where $n_e=Zn_i$ is the electron 
density, $p_F$ -- the Fermi momentum, and  
$m_e$ -- the electron (vacuum) mass.

\section{Thermoelectric transport from the Boltzmann equation}
\label{sec:Boltzmann}

\subsection{Solving the Boltzmann equation}
\label{sec:solve_Boltzmann}

In this work, we mainly follow Refs.~\cite{Harutyunyan2016,Harutyunyan2024-2} to present the Boltzmann framework for thermoelectric transport in the presence of magnetic fields. The electron dynamics is governed by the Boltzmann equation for the electron distribution function
%-----------------------------------
\bea\label{eq:boltzmann}
\frac{\partial f}{\partial t}+
\bm v\frac{\partial f}{\partial\bm r}-
e(\bm E+\bm v\times \bm B)\frac{\partial f}
{\partial\bm p}=I,
\eea
%-----------------------------------
where $\vecE$ and $\vecB$ are the electric and magnetic fields, 
$\bm v =\partial\ep/\partial \bm p$ is the electron velocity with
$\ep=\sqrt{p^2+m_e^2}$, $e$ is the unit charge, and $I = I_{ei} + I_{en}$ is the collision integral which includes the electron-ion and electron-neutron scattering processes. These integrals are given by
%-----------------------------------
\bea\label{eq:collision_ei}
I_{ei}&=&-(2\pi)^4\sum\limits_{234}|{\cal M}^{ei}_{12\to 34}|^2 
\delta^{(4)}(p_1+p_2-p_3-p_4)\nonumber\\
&& \times\big[f_1(1-f_3)g_2-f_3(1-f_1)g_4\big],\\
\label{eq:collision_en}
I_{en} &=&-(2\pi)^4\sum\limits_{234}|{\cal M}^{en}_{12\to 34}|^2 
\delta^{(4)}(p_1+p_2-p_3-p_4)\nonumber\\
&& \times
\big[f_1 (1-f_3) f'_2(1-f'_4) - f_3 (1-f_1) f'_4(1-f'_2)\big],\nonumber\\
\eea
%-----------------------------------
where (using short-hand notation) $f_{1,3}\equiv f(p_{1,3})$ and
$f'_{2,4}\equiv f'(p_{2,4})$ are the distribution functions of electrons and
neutrons, $g_{2,4}\equiv g(p_{2,4})$ is the distribution function of ions,
${\cal M}^{ei}_{12\to 34}$ and ${\cal M}^{en}_{12\to 34}$ are the
electron-ion and electron-neutron scattering matrix elements,
respectively, and $\sum\limits_{i}\equiv \int d\bm p_i/(2\pi)^3$.  We will
neglect the electron-electron collisions, as their rate is
suppressed by a factor of $Z^{-1}$ as compared to the rate of
electron-ion scattering.  As ions form a classical fluid in
equilibrium, the function $g(p)$ is given by the Maxwell-Boltzmann
distribution
%-----------------------------------
\bea\label{eq:maxwell}
g(p)=n_i\bigg(\frac{2\pi}{MT}\bigg)^{3/2}
\exp\left(-\frac{p^2}{2MT}\right).
\eea
%-----------------------------------
Additionally, we assume that the neutrons remain in thermal
equilibrium and are therefore described by the Fermi–Dirac
distribution
%-----------------------------------
\bea f(\epsilon)=\frac{1}{e^{\beta(\epsilon-\mu_n)}+1},
\qquad \epsilon = \frac{p^{2}}{2m_n^*},
\eea
%-----------------------------------
where $m_n^*$ and $\mu_n$ are the effective neutron mass and chemical potential, respectively. In principle, the neutron effective mass should be computed self-consistently within the microscopic model that determines the composition of matter. However, its value is expected to deviate only weakly from the vacuum mass, and in what follows we therefore take $m_n^* \simeq m_n$.

To linearize the kinetic equation~\eqref{eq:boltzmann}, the electron distribution is written as
%-----------------------------------
\bea\label{eq:distribution}
f= f^0+\delta f,\quad \delta f=-\phi
\frac{\partial f^0}{\partial\ep},
\eea
%-----------------------------------
where $f^0(\ep)$ is the equilibrium Fermi-Dirac distribution function
of electrons and $\delta f\ll f^0$ is a small perturbation. To derive
the thermal conductivity under stationary conditions, we assume that
the temperature $T=\beta^{-1}$ and the electron chemical potential
$\mu$ vary slowly in space, i.e., $T=T(\bm r)$, $\mu=\mu(\bm r)$ and
are independent of time. The leading-order non-vanishing derivatives 
appearing in the Boltzmann equation \eqref{eq:boltzmann} then can be written as 
%-----------------------------------
\bea\label{eq:fermi_deriv_r}
\frac{\partial f^0}{\partial \bm r}&=&
\frac{\partial f^0}{\partial T}
\frac{\partial T}{\partial \bm r}
+\frac{\partial f^0}{\partial \mu}
\frac{\partial \mu}{\partial \bm r}\nonumber\\
&=& \frac{\partial f^0}{\partial \ep}
\left(\frac{\mu -\ep}{T}
\bm{\nabla} T-\bm{\nabla} \mu\right),\\
\label{eq:fermi_deriv_p}
\frac{\partial f^0}{\partial\bm p} &=&
\bm v\frac{\partial f^0}{\partial\ep},
\quad\frac{\partial f^0}{\partial\ep}=
-\beta f^0(1-f^0).
\eea
%-----------------------------------
Note that in all terms involving spatial derivatives or the electric field, one may substitute 
$f\to f^0$. However, in the magnetic term, the perturbation must be retained, since in equilibrium the magnetic-field contribution vanishes due to
$[\vecv\times \vecB]({\partial f^0}/{\partial\vecp})\propto[\vecv\times \vecB]\vecv=0.$ Therefore, for
the third term in Eq.~\eqref{eq:boltzmann} we obtain [using
Eqs.~\eqref{eq:distribution} and \eqref{eq:fermi_deriv_p}]
%-----------------------------------
\bea\label{eq:field_term}
e(\bm E+\bm v\times \vecB)\frac{\partial f}
{\partial\bm p}=e\bm v\cdot\bm E\frac{\partial f^0}
{\partial\ep}-e[\bm v\times\vecB]
\frac{\partial f^0}{\partial\ep}
\frac{\partial\phi}{\partial\bm p}.
\eea
%-----------------------------------
Substituting Eqs.~\eqref{eq:fermi_deriv_r}--\eqref{eq:field_term} in
Eq.~\eqref{eq:boltzmann}, at linear order in macroscopic gradients, we
find
%-----------------------------------
\bea\label{eq:boltzmann_linear}
\frac{\partial f^0}{\partial \ep}
\left(e\bm v\cdot\bm F
-e[\bm v\times\vecB]\frac{\partial\phi}{\partial\bm p}\right)=-I[\phi],
\eea
%-----------------------------------
where we defined
%-----------------------------------
\bea\label{eq:vector_F}
\bm F=\bm E'+\frac{\ep-\mu}{eT}\bm{\nabla} T,
\qquad e\bm E'=e\bm E+\bm{\nabla} \mu.
\eea
%-----------------------------------
Linearized collision integrals are given by
%-----------------------------------
\bea\label{eq:coll_ei_lin}
I_{ei}[\phi] &=&-(2\pi)^4\beta\sum\limits_{234}
|{\cal M}^{ei}_{12\to 34}|^2
\delta^{(4)}(p_1+p_2-p_3-p_4)\nonumber\\
&&f^0_1(1-f^0_3)g_2(\phi_1-\phi_3),\\
\label{eq:coll_en_lin}
I_{en}[\phi]  
&=& -(2\pi)^4\beta\sum\limits_{234}
|{\cal M}^{en}_{12\to 34}|^2
\delta^{(4)}(p_1+p_2-p_3-p_4)\nonumber\\
&&f_1^{0}(1 - f_3^{0})f'_2(1 - f'_4)(\phi_1 - \phi_3),\qquad
\eea
%-----------------------------------
where, as usual, we used the fact that the collision integral vanishes
in equilibrium, since the corresponding combinations of distribution
functions vanish.  We search the solution of
Eq.~\eqref{eq:boltzmann_linear} in the form
%-----------------------------------
\bea\label{eq:solution_cond}
\phi =\bm p\cdot\bm A(\ep),
\eea
%-----------------------------------
where the vector $\bm A$ depends only on the absolute value of $\bm p$.
From Eqs.~\eqref{eq:boltzmann_linear}--\eqref{eq:solution_cond} we have
%-----------------------------------
\bea
\label{eq:boltzmann_cond2}
e\bm v\frac{\partial f^0}{\partial\ep}
(\bm F+[\bm A\times \bm B])=-I_{ei}-I_{en},
\eea
%-----------------------------------
with 
%-----------------------------------
\bea\label{eq:coll_ei_lin1}
I_{ei} &=& -(2\pi)^4\beta\sum\limits_{234}|{\cal M}^{ei}_{12\to 34}|^2
\delta^{(4)}(p_1+p_2-p_3-p_4)\nonumber\\
&\times&
f^0_1(1-f^0_3)g_2
(\bm p_1\cdot\bm A_1-\bm p_3\cdot\bm A_3),\\
\label{eq:coll_en_lin1}
I_{en} &=& -(2\pi)^4\beta\sum\limits_{234}|{\cal M}^{en}_{12\to 34}|^2
\delta^{(4)}(p_1+p_2-p_3-p_4)\nonumber\\
&\times&
f_1^{0}(1-f_3^{0}) f'_2(1-f'_4)
(\bm p_1\cdot\bm A_1-\bm p_3\cdot\bm A_3).
\eea
% -----------------------------------
Since the energy transfer in collisions with both nuclei and neutrons
is small, we approximate
$\bm A_1 \approx \bm A_3$ in Eqs.~\eqref{eq:coll_ei_lin1} and
\eqref{eq:coll_en_lin1}, allowing
$\bm A_1$ to be taken outside the collision integrals $(\bm p_1 \equiv
\bm p)$
%-----------------------------------
\bea\label{eq:collision_cond1}
I_{ei} &=& -(2\pi)^4\beta\sum\limits_{234}
|{\cal M}^{ei}_{12\to 34}|^2\delta^{(4)}
(p_1+p_2-p_3-p_4)\nonumber\\
&&\hspace{-1.cm}\times f^0_1(1-f^0_3)g_2[\bm A\cdot (\bm p_1-\bm p_3)]
=\frac{\partial f^0}{\partial\ep}
(\bm A\cdot\bm p)\,\tau_{ei}^{-1}(\ep),\\
\label{eq:collision_cond1_en}
I_{en} &=& -(2\pi)^4\beta\sum\limits_{234}
|{\cal M}^{en}_{12\to 34}|^2\delta^{(4)}
(p_1+p_2-p_3-p_4)\nonumber\\
&&\hspace{-1.2cm} \times f_1^{0}(1-f_3^{0}) f'_2(1-f'_4)[\bm A\cdot(\bm p_1-\bm p_3)]
=\frac{\partial f^0}{\partial\ep}
(\bm A\cdot\bm p)\,\tau_{en}^{-1}(\ep),\nonumber\\
\eea
%-----------------------------------
where $\bm q =\bm p_1-\bm p_3=\bm p_4-\bm p_2$ is the transferred momentum, and we defined electron-ion and electron-neutron collision rates, \ie, inverse relaxation times by ($\ep\equiv\ep_1$)
%-----------------------------------
\bea\label{eq:t_relax}
\tau_{ei}^{-1}(\ep) &=& (2\pi)^{-5}\!
\int\! d\bm q\!\int\! d\bm p_2\,
|{\cal M}^{ei}_{12\to 34}|^2 
\frac{\bm q\cdot \bm p}{p^2}\nonumber\\ 
&& 
\delta(\ep+\ep_2-\ep_3-\ep_4)
\frac{1-f^0_3}{1-f^0_1} g_2,\\\
\label{eq:t_relax_en}
\tau_{en}^{-1}(\ep) &=& (2\pi)^{-5}\!
\int\! d\bm q\!\int\! d\bm p_2\,
|{\cal M}^{en}_{12\to 34}|^2 
\frac{\bm q\cdot \bm p}{p^2}\nonumber\\ 
&& 
\delta(\ep+\ep_2-\ep_3-\ep_4)
\frac{1-f^0_3}{1-f_1^{0}}f'_2(1-f'_4).
\eea
%-----------------------------------
Substituting the collision integrals given by Eqs.~\eqref{eq:collision_cond1} and \eqref{eq:collision_cond1_en}
into Eq.~\eqref{eq:boltzmann_cond2} we obtain
%-----------------------------------
\bea\label{eq:boltzmann_cond3}
e\tau\ep^{-1}\bm F+\omega_c
\tau[\bm A\times\bm b]+\bm A=0,
\eea
%-----------------------------------
where 
%-----------------------------------
\bea
\tau^{-1} = \tau_{en}^{-1} + \tau_{ei}^{-1}
\eea
% -----------------------------------
is the effective electron relaxation rate,
$\omega_c=eB\ep^{-1}$
is the cyclotron frequency, and $\bm b=\bm B/B$ is the 
unit vector along the magnetic field.
We search the vector $\bm A$ in the form
$\bm A=\alpha\bm f+\beta (\bm f\cdot\bm b)
\bm b+\gamma[\bm f\times\bm b]$, with unit vector
$\bm f=\bm F/F$ and 
energy-dependent coefficients $\alpha$, $\beta$,
$\gamma$. Substituting this
expression into Eq.~\eqref{eq:boltzmann_cond3}
and equating the coefficients of the three independent vectors 
$\bm f$, $\bm b$ and $[\bm f\times\bm b]$, we obtain
%-----------------------------------
\bea\label{eq:psi_cond}
\phi = -\frac{e\tau}{1+(\omega_c\tau)^2}
v_i\left[\delta_{ij}-\omega_c\tau\varepsilon_{ijk}
b_k+(\omega_c\tau)^2b_i b_j\right]F_j.\nonumber\\
\eea
%-----------------------------------
The electron–ion and electron–neutron scattering matrix elements can
be evaluated using standard QED techniques for thermal media. The
electron–ion scattering amplitude incorporates ion–ion correlations
through the structure factor $S(q)$ of an OCP.  For numerical
computations, we will use the structure factor of
OCP~\cite{Galam1976,Itoh1983,Tamashiro1999}, previously employed in
Refs.~\cite{Harutyunyan2016,Harutyunyan2024-1,Harutyunyan2024-2}. 
In  using $S(q)$ we make the static, quasi-elastic approximation to the
ionic response. A fully dynamical treatment would replace this input by
$S(q,\omega)$ in the collision integral. The small energy transfer is
kinematically supported by the large ion mass: a typical ion thermal
velocity is $v_i\sim\sqrt{T/M_i}$, which is of order $10^{-2}$ or less
for heavy inner-crust nuclei even at $T=10$~MeV. Hence the static
approximation captures the leading heavy-ion limit. This observation,
however, does not provide a universal percent-level error estimate for
the correlation contribution. Schematically, if a more complete ionic
model changes the static correlations by $\delta S(q)$, the relative
change of the electron-ion rate is of the order
$$\int dq\,W(q)\delta S(q)\left[\int dq\,W(q)S(q)\right]^{-1},$$ with $W(q)$ denoting
the remaining kernel in Eq.~\eqref{eq:relax_time3}. Recent
molecular--dynamics calculations of warm neutron-star crust matter, including ion
motion, electron screening, and finite-size ions, illustrate that the
ionic thermodynamics can change appreciably in the warm, dense regime
~\cite{BarbaGonzalez2026}; they do not by themselves provide a direct
correction to the electron transport rate considered here. A
quantitative transport estimate of this correction would require a
consistent $S(q,\omega)$ calculation and is beyond the scope of the
present work.

In the present work, electron–ion scattering in the inner crust is
treated assuming spherical nuclei with a uniform proton charge
distribution, and non-spherical (“pasta”) phases are not
included. Finite nuclear size effects are incorporated through the
single-ion charge form factor~\cite{1984ApJ...285..758I}
%-----------------------------------
\bea\label{eq:formfactor}
F(q)=-3\frac{qr_{c}\cos(qr_{c})-\sin(qr_{c})}{(qr_{c})^3},
\eea 
%-----------------------------------------------------------
which corresponds to a uniformly charged sphere and is intended to account for finite nuclear size effects at the level commonly adopted in neutron star crust transport calculations. 

With these approximations, the electron-ion relaxation is given by the
well-known formula (see Ref.~\cite{Harutyunyan2016} and references
therein)
%-----------------------------------
\bea\label{eq:relax_time3}
\tau^{-1}_{ei}(\ep) = \frac{\pi Z^2\alpha^{2}n_i}{\ep \, p^3}
\int_{0}^{2p} dq\, q^3 S(q) F^2(q)\frac{4\ep^2
  -q^2}{|q^2+q_D^2|^2},
\eea 
%-----------------------------------
where the Debye screening is treated via the low-frequency limit of the HTL polarization tensor of a QED plasma (detailed in Ref.~\cite{Harutyunyan2016} Sec.~IV), $\alpha=1/137$ is the fine-structure constant, and the Debye wave-number $q_D$ is defined as
%-----------------------------------
\bea\label{eq:Debye}
q_D^{2} =-\frac{4\alpha}{\pi}\!\int_0^{\infty}\!
dp\,p^2\frac{\partial f^0}{\partial\ep}.
\eea
%-----------------------------------

  It is useful to distinguish the three ingredients entering
  Eq.~\eqref{eq:relax_time3}. The factor $F(q)$ is a
  \emph{single-nucleus charge form factor}: it resolves the finite
  charge distribution of a uniform spherical nucleus in the
  electron--ion scattering vertex. The factor $q_D$ describes
  electronic screening of that vertex. In contrast, $S(q)$ describes
  correlations among ion centers and is taken from a point-ion Coulomb
  OCP. Thus $F(q)$ does not account for changes of $S(q)$, the ionic
  equation of state, or the crystallization condition that arise when
  the ion--ion interaction itself is modified by screening or finite
  charge extent; screened (Yukawa) fluids are a simple example in
  which the thermodynamics and fluid--solid boundary depend on the
  screening strength~\cite{KhrapakThomas2015}. Furthermore, $F(q)$
  does not include diffuse proton density profiles, overlap of
  extended charge distributions, deformation of clusters, or pasta
  topology.

These omissions become progressively more relevant in the deepest
layers of the inner crust, where the nuclear radius can approach the
Wigner--Seitz cell radius, as explicitly indicated by
Fig.~\ref{fig:radii}. In this sense our treatment is the leading
finite-size correction to an OCP transport calculation: finite nuclear
size is kept in the electron scattering amplitude, whereas its
feedback on ionic correlations is not. For an actual multi-component
plasma the single OCP factor $Z^2S(q)$ must moreover be replaced by
the appropriate composition-weighted correlations (or, in a fully
dynamical calculation, by partial dynamical structure factors). An
impurity parameter can be a useful characterization near
crystallization~\cite{Carreau2020MCP}, but a broad liquid composition
generally requires the full species distribution and cross
correlations. We therefore regard the present coefficients as a
single-component reference study rather than a quantitative treatment
of the most compositionally complex accreting or merger matter. Such
an MCP extension is beyond the scope of the present work.

The scattering matrix for the electron scattering off the anomalous
magnetic moment of neutrons at finite temperatures is computed in
Appendix~\ref{app:matrix_en}, which is followed by the computation of
the corresponding collision rate in Appendix~\ref{app:relax_en}.  We
find that in the regime of interest $\tau_{en} \gg \tau_{ei}$,
indicating that electron–neutron scattering has a negligible effect on
electron transport in the liquid phase in the inner crust across the
temperatures considered. The minor role of electron–neutron scattering
in the inner crust was also noted earlier in Ref.~\cite{Flowers1976},
which focused on low-temperature crystallized matter, a conclusion
that is also valid if the induced interaction between electrons and
neutrons is taken into account~\cite{Bertoni2015}.

\subsection{Thermoelectric currents and transport coefficients}

Now we are in a position to calculate the electrical and thermal
currents using the solution~\eqref{eq:distribution},
\eqref{eq:psi_cond} and the expression \eqref{eq:vector_F} for $\bm F$
%-----------------------------------
\bea\label{eq:jk}
j_k &=& -\int\frac{2d\bm p}{(2\pi)^3}ev_k\delta f 
= \sigma_{kj}E'_j-\alpha_{kj}\partial_j T,\\
\label{eq:qk}
q_k &=&\int\frac{2d\bm p}{(2\pi)^3}(\ep-\mu)v_k\delta f 
=\tilde{\alpha}_{kj}E'_j
-\tilde{\kappa}_{kj}\partial_j T,
\eea
%-----------------------------------
where the transport coefficient tensors are defined as
%-----------------------------------
\bea
\sigma_{kj} & = & -\int\frac{2d\bm p}{(2\pi)^3}
\frac{\partial f^0}{\partial\ep}
\frac{e^2\tau}{1+(\omega_c\tau)^2}
v_kv_i\nonumber\\
&\times&\left[\delta_{ij}-\omega_c\tau\ep_{ijm}
b_m +(\omega_c\tau)^2b_ib_j\right], \\
\alpha_{kj} & = & \int\frac{2d\bm p}{(2\pi)^3}
\frac{\partial f^0}{\partial\ep}
\frac{e(\ep-\mu)\tau}{1+(\omega_c\tau)^2}
v_kv_i\nonumber\\
&\times&\left[\delta_{ij}-\omega_c\tau\ep_{ijm}
b_m +(\omega_c\tau)^2b_ib_j\right]T^{-1}, \\
\tilde{\alpha}_{kj}& = &\int\frac{2d\bm p}{(2\pi)^3}
\frac{\partial f^0}{\partial\ep}
\frac{e(\ep-\mu)\tau}{1+(\omega_c\tau)^2}
v_kv_i\nonumber\\
&\times&\left[\delta_{ij}-\omega_c\tau\ep_{ijm}
b_m +(\omega_c\tau)^2b_ib_j\right], \\
\tilde{\kappa}_{kj}& = & -\int\frac{2d\bm p}{(2\pi)^3}
\frac{\partial f^0}{\partial\ep}
\frac{(\ep-\mu)^2\tau}{1+(\omega_c\tau)^2}
v_kv_i\nonumber\\
&\times&\left[\delta_{ij}-\omega_c\tau\ep_{ijm}
b_m +(\omega_c\tau)^2b_ib_j\right]T^{-1}.
\eea
% -----------------------------------
Equations~\eqref{eq:jk} and \eqref{eq:qk} can equivalently be written
as
%-----------------------------------
\bea\label{eq:currents_reversed}
\bm E' = \hat{\varrho}\bm j 
-\hat{Q}\bm{\nabla} T,\qquad
\bm q = -\hat{\kappa}\bm{\nabla} T
-T\hat{Q}\bm j,
\eea
%-----------------------------------
where $\hat{\sigma}$ and $\hat{\varrho}=\hat{\sigma}^{-1}$ are the
conductivity and resistivity tensors, respectively;
$\hat{Q}=-\hat{\varrho}\hat{\alpha} $ is the thermopower and
$\hat{\kappa} =\hat{\tilde{\kappa}}+ T\hat{\alpha} \hat{Q} $ is the
thermal conductivity. The tensors $\hat{\sigma}$, $\hat{\alpha}$,
$\hat{\tilde{\alpha}}$ and $\hat{\tilde{\kappa}}$ are given by
%-----------------------------------
\bea\label{eq:cond_tensors}
\hat{\sigma} =e^2{\hat{\cal L}}^{0},\qquad
\hat{\tilde\alpha}=T\hat{\alpha} =-eT{\hat{\cal L}}^{1},\qquad
\hat{\tilde{\kappa}}=T\hat{\cal L}^{2},
\eea
%-----------------------------------
with 
%-----------------------------------
\bea\label{eq:L_tensors}
{\cal L}^{n}_{kj}&=&\delta_{kj}{\cal L}^{n}_{0}-\varepsilon_{kjm}
b_m{\cal L}^{n}_{1} +b_kb_j{\cal L}^{n}_{2},\\
\label{eq:L_components}
{\cal L}^{n}_{l}&=& -\frac{1}{3\pi^2}\int_{m_e}^\infty d\ep\frac{\partial f^0}{\partial\ep}
\left(\frac{\ep-\mu}{T}\right)^n
{\cal F}_l(\ep),\nonumber\\
{\cal F}_l(\ep)&=&\frac{p^3}{\ep}
\frac{\tau(\omega_c\tau)^l}{1+(\omega_c\tau)^2},\quad l=0,1,2.
\eea
%-----------------------------------
In the absence of a magnetic field 
$\hat{\cal L}^{n}$ tensors become diagonal ${\cal L}^{n}_{kj}=\delta_{kj}{\cal L}^{n}$ with
%-----------------------------------
\bea\label{eq:L_parallel}
{\cal L}^{n}=-\frac{1}{3\pi^2}\int_{m_e}^\infty\! d\ep\, \frac{p^3}{\ep}
\frac{\partial f^0}{\partial\ep}
\left(\frac{\ep-\mu}{T}\right)^n\tau
={\cal L}^{n}_{0}+{\cal L}^{n}_{2}.
\eea
%-----------------------------------
With the magnetic field along $z$ axis, we have
%-----------------------------------
\bea\label{eq:L_matrix}
\hat{\cal L}^{n}=
\begin{pmatrix}
{\cal L}^{n}_0 & -{\cal L}^{n}_1 & 0 \\
{\cal L}^{n}_1 & {\cal L}^{n}_0 & 0 \\
0 & 0 & {\cal L}^{n}
\end{pmatrix},
\eea
%-----------------------------------
where ${\cal L}^{n}$, ${\cal L}^{n}_0$ and ${\cal L}^{n}_1$ are,
respectively, the longitudinal, transverse, and Hall components of this
tensor.  A similar tensor structure applies to all transport
coefficients defined above. Using
Eqs.~\eqref{eq:cond_tensors}--\eqref{eq:L_matrix}, it is
straightforward to obtain the components of thermopower
%-----------------------------------
\bea\label{eq:Q_comp}
Q =-\frac{\alpha}{\sigma},~
Q_0 =-\frac{\alpha_0\sigma_0 + \alpha_1\sigma_1}{\sigma_0^2+\sigma_1^2},~
Q_1 = -\frac{\alpha_1\sigma_0
  -\alpha_0\sigma_1}{\sigma_0^2+\sigma_1^2}, \nonumber\\
\eea
%-----------------------------------
and thermal conductivity
%-----------------------------------
\bea\label{eq:kappa_scalar}
\kappa &=& \tilde{\kappa}+T\alpha Q
=\tilde{\kappa}-T\frac{\alpha^2}{\sigma},\\
\label{eq:kappa_0}
\kappa_0 &=& \tilde{\kappa}_0+
T\big(\alpha_0 Q_0 - \alpha_1 Q_1\big),\\
\label{eq:kappa_1}
\kappa_1 &=& \tilde{\kappa}_1
+T\big(\alpha_0 Q_1 +\alpha_1 Q_0\big).
\eea
%-----------------------------------
Thus, the components of conductivities and thermopower are fully
determined if the relaxation time $\tau$ is known. The low-temperature
expressions of the transport coefficients are derived in
Appendix~\ref{app:low-T}.

\section{Numerical results}
\label{sec:results}

Numerically, the thermal conductivity and thermopower are evaluated
using the relaxation time given by Eq.~\eqref{eq:relax_time3} and the
formulas~\eqref{eq:cond_tensors},
\eqref{eq:Q_comp}--\eqref{eq:kappa_1}.  We recall that for large
magnetic fields ($\omega_c\tau \gtrsim 1$) the full tensor structure
of these coefficients must be taken into account, while for weak
magnetic fields ($\omega_c\tau \ll 1$) only the longitudinal
components $\kappa$ and $Q$ are relevant. Below, we will study the
dependence of these transport coefficients on density, temperature,
and magnetic field strength for the selected compositions. The numerical
results will be presented in c.g.s. units, which are common in
astrophysical applications.

\subsection{Relaxation time and Hall parameter}

%-------------------------------------------------------
\begin{figure}[hbt] 
\begin{center}
\includegraphics[width=\linewidth,keepaspectratio]{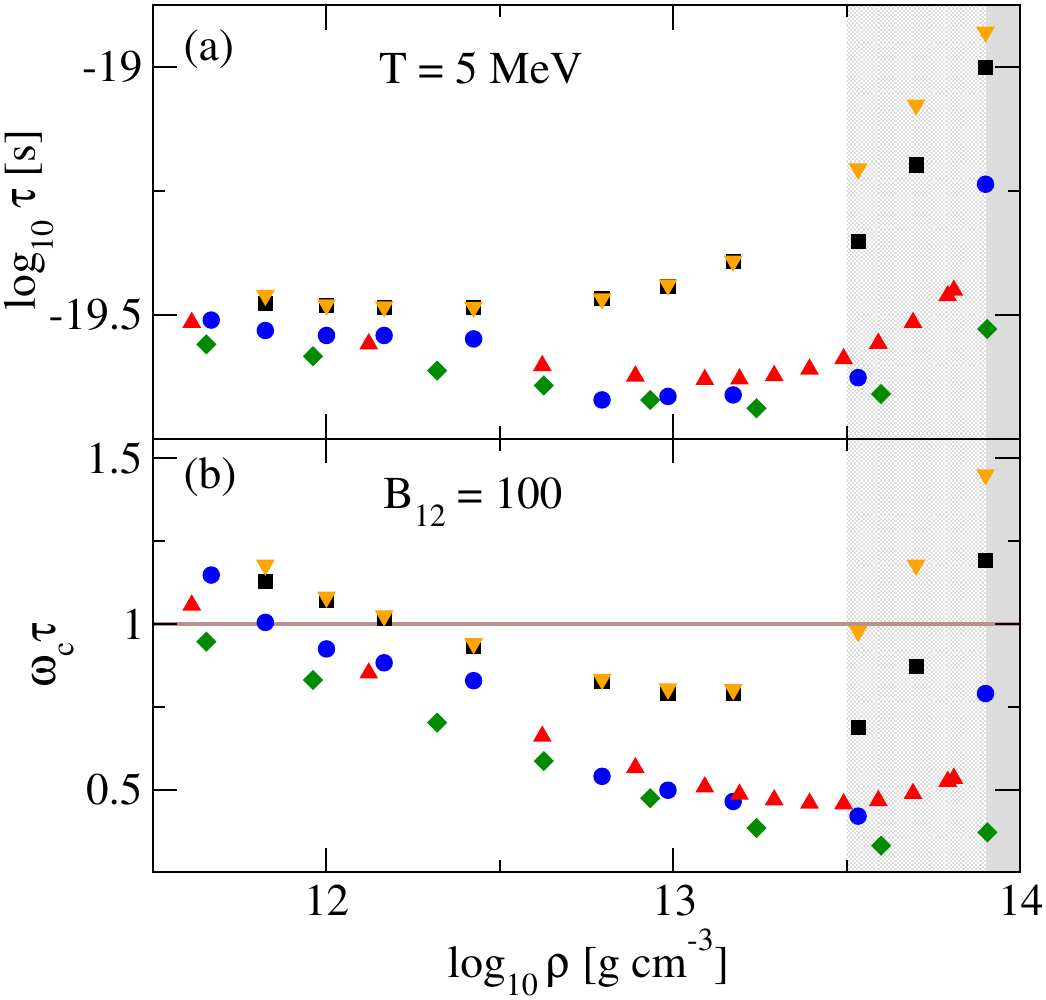}
\caption{ (a) The relaxation time $\tau$ and (b) the Hall parameter
  $\omega_c\tau$ as functions of density for five compositions as
  labeled in Fig.~\ref{fig:radii}. The temperature is fixed at
  $T=5$~MeV, and the magnetic field is fixed at $B_{12}=100$.  The
  shaded regions are the same as in Fig.~\ref{fig:radii}.
  }
\label{fig:tau_dens}
\end{center}
\end{figure}
%-------------------------------------------------------
\begin{figure}[hbt] 
\begin{center}
\includegraphics[width=\linewidth,keepaspectratio]{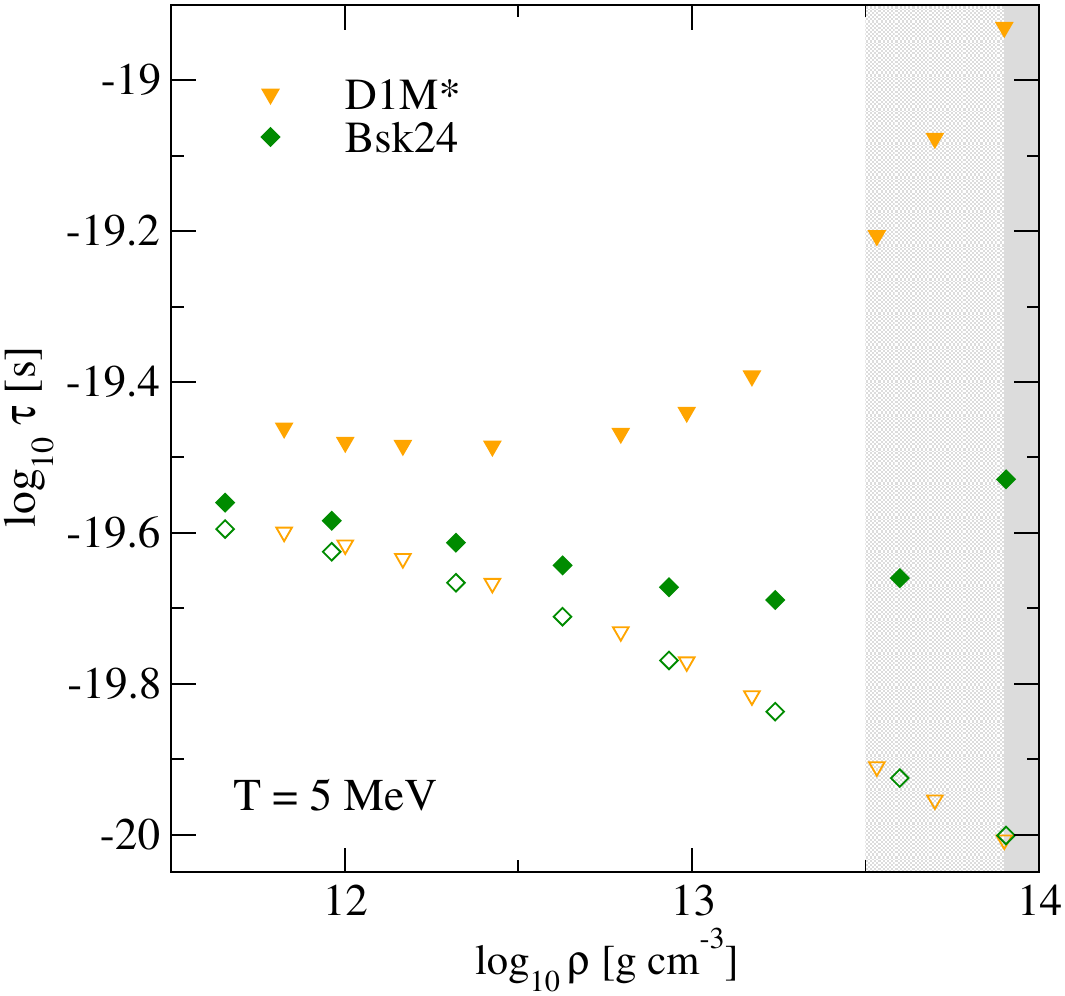}
\caption{Density-dependence of the relaxation time for two models
  evaluated with full nuclear form factor $F(q)$ (filled symbols) and
  with $F(q)=1$ (empty symbols). The temperature is fixed at
  $T=5$~MeV.
  The shaded regions are the same as in Fig.~\ref{fig:radii}.
  }
\label{fig:tau_noform}
\end{center}
\end{figure}
%-------------------------------------------------------

The behavior of the relaxation time, $\tau$, and the Hall parameter,
$\omega_c \tau$, was discussed in detail in
Ref.~\cite{Harutyunyan2024-1} in the context of electrical
conductivity; here, we summarize the most important features.
Figure~\ref{fig:tau_dens} shows the density dependence of the
relaxation time and the Hall parameter at the Fermi energy for five
different compositions at $T = 5$~MeV. The variations in $\tau$ and
$\omega_c \tau$ among the compositions are primarily due to
differences in nuclear size, which affect the nuclear form factor,
$F(q)$, particularly at high densities, $\log_{10} \rho$ [g cm$^{-3}$]
$> 13$.

The finite nuclear size leads to an increase in the relaxation time with density, due to the suppression of electron–ion scattering rates compared to scattering off a point-like nucleus. This effect is particularly significant for the D1M and D1M* models, which have larger nuclei in the high-density regime, as seen in Fig.~\ref{fig:radii}. Figure~\ref{fig:tau_noform} shows the relaxation time for the D1M* and Bsk24 models, which predict very similar values of $Z$ and $R_{\rm WS}$, but rather different values of $A$ and $r_c$. Consequently, the effect of the nuclear form factor is markedly different for these two models. Indeed, as seen in the figure, the relaxation times for both models would be nearly identical if the effect of $F(q)$ were neglected, i.e., $F(q)=1$ (empty symbols).

Panel (b) of Fig.~\ref{fig:tau_dens} shows the Hall parameter for a magnetic field $B_{12} \equiv B/(10^{12}\,{\rm G}) = 100$. At this field strength, $\omega_c \tau$ is of order unity throughout the inner crust. Consequently, the inner crust becomes anisotropic at such field strengths, although the effect is already negligible for $B_{12} \lesssim 13$. We also note that the anisotropy is more pronounced at higher densities for models with larger nuclei, where the nuclear form factor plays a dominant role.

\subsection{Longitudinal thermal conductivity} 

%-------------------------------------------------------
\begin{figure}[t] 
\begin{center}
\includegraphics[width=\linewidth,keepaspectratio]{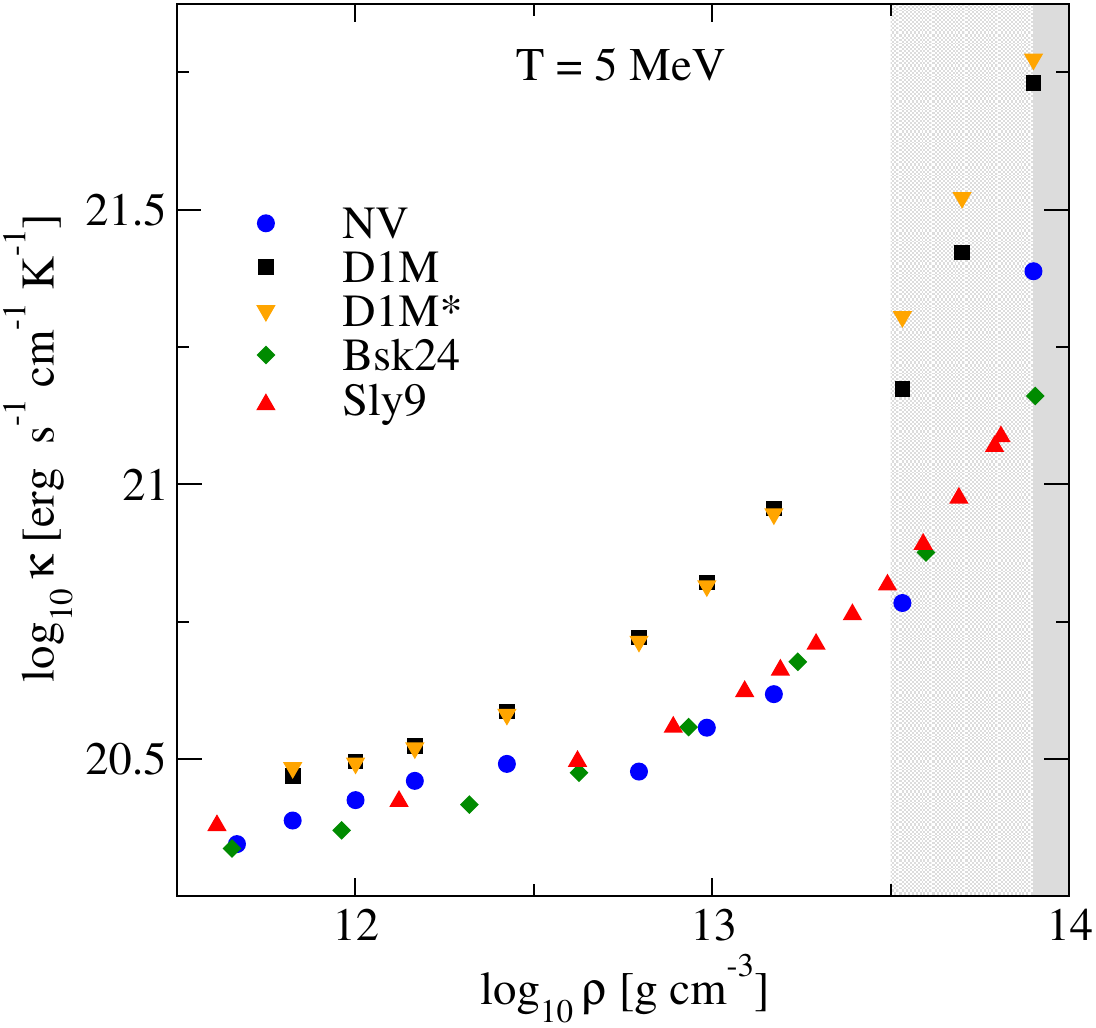}
\caption{Dependence of the scalar conductivity on density for five
  compositions. The temperature is fixed at $T=5$~MeV.  The shaded
  regions are the same as in Fig.~\ref{fig:radii}.  }
\label{fig:kappa_dens}
\end{center}
\end{figure}
%-------------------------------------------------------
\begin{figure}[hbt]
  \begin{center}
\includegraphics[width=\linewidth,keepaspectratio]{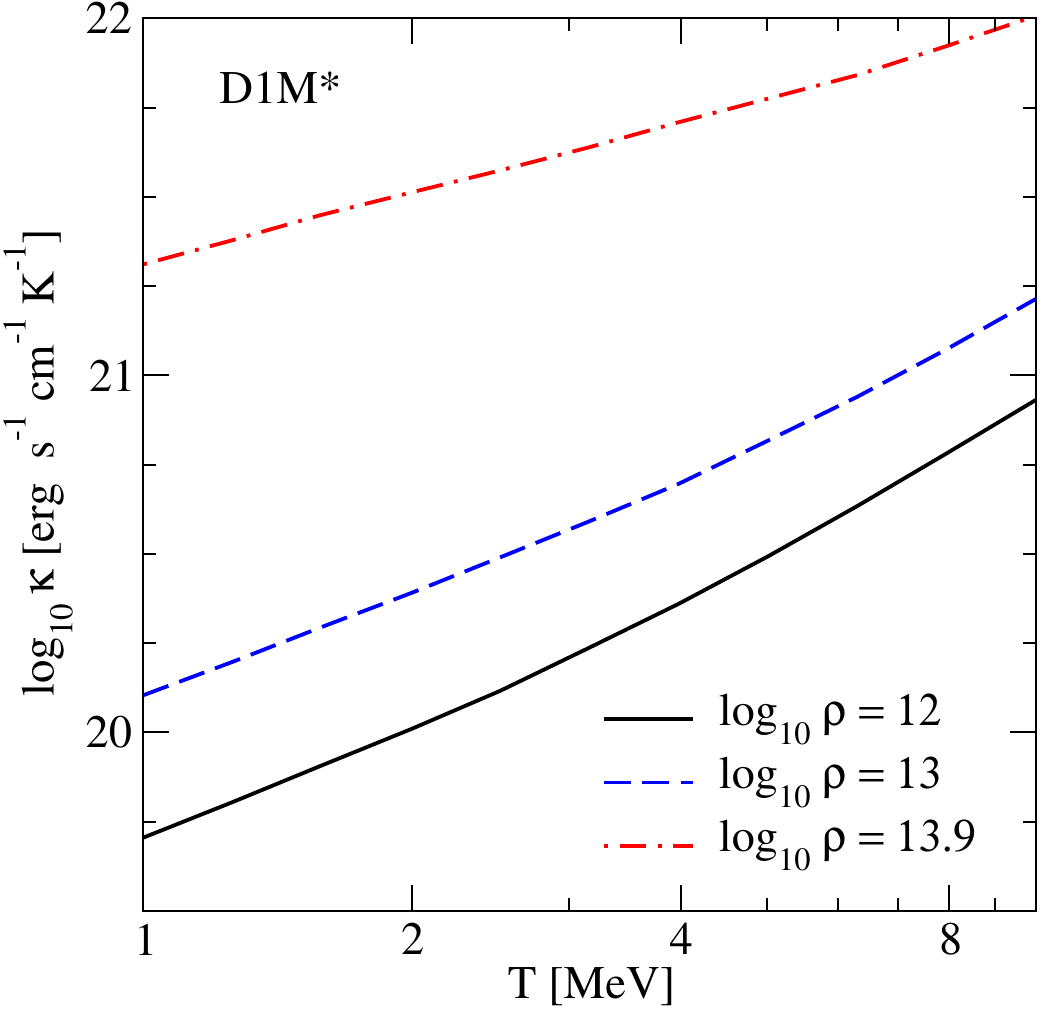}
\caption{ The temperature dependence of the scalar 
conductivity for various values of the density for composition D1M*.}.
\label{fig:kappa_temp}
\end{center}
\end{figure}
% --------------------------------------------------------
%------------------------------------------------------
\begin{figure}[hbt] 
\begin{center}
  \includegraphics[width=0.95\linewidth,keepaspectratio]{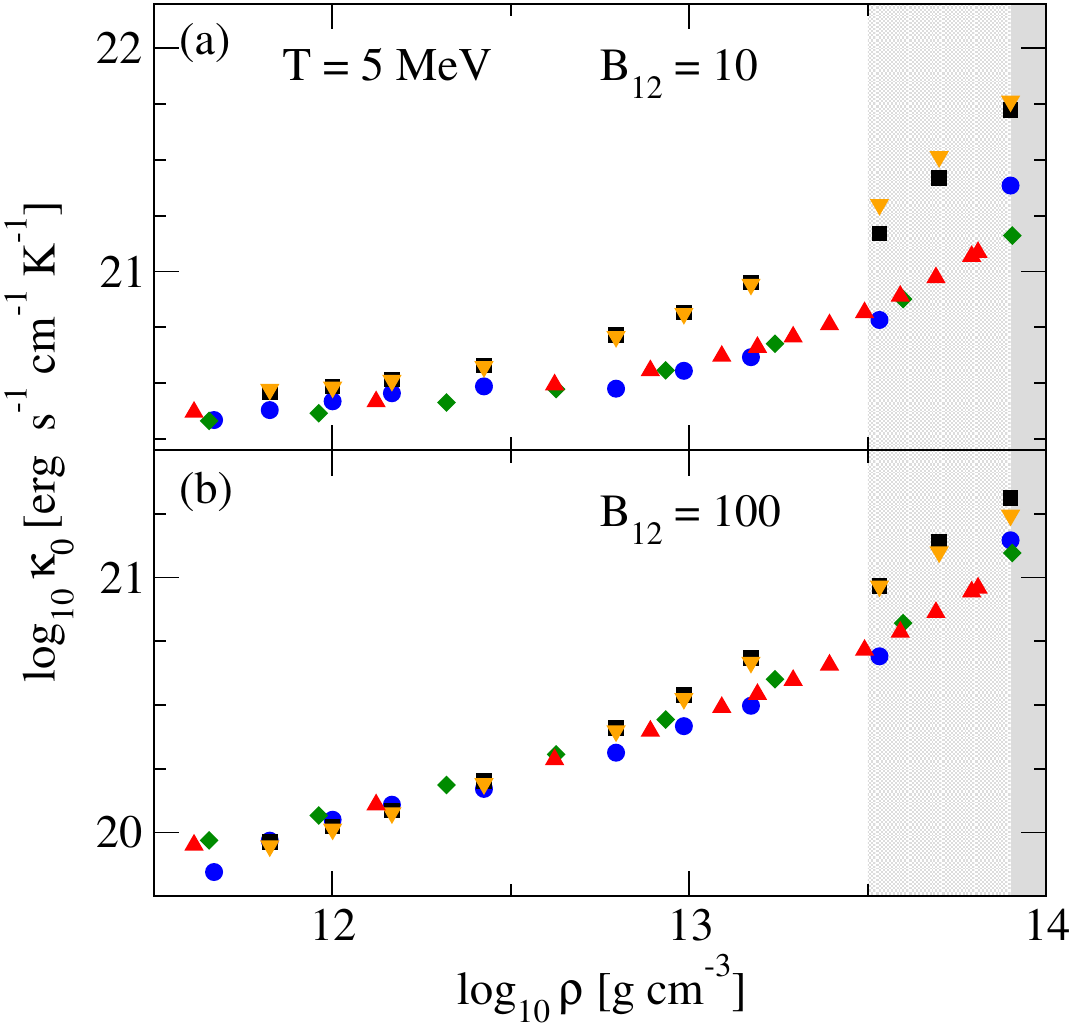}
  \caption{The same as in Fig.~\ref{fig:kappa_dens} but for $\kappa_0$
    and for two magnetic fields $B_{12}=10$ and $B_{12}=100$. The
    shaded regions are the same as in Fig.~\ref{fig:radii}.  }
\label{fig:kappa0_dens}
\end{center}
\end{figure}
%------------------------------------------------------
\begin{figure}[hbt] 
\begin{center}
  \includegraphics[width=0.95\linewidth,keepaspectratio]{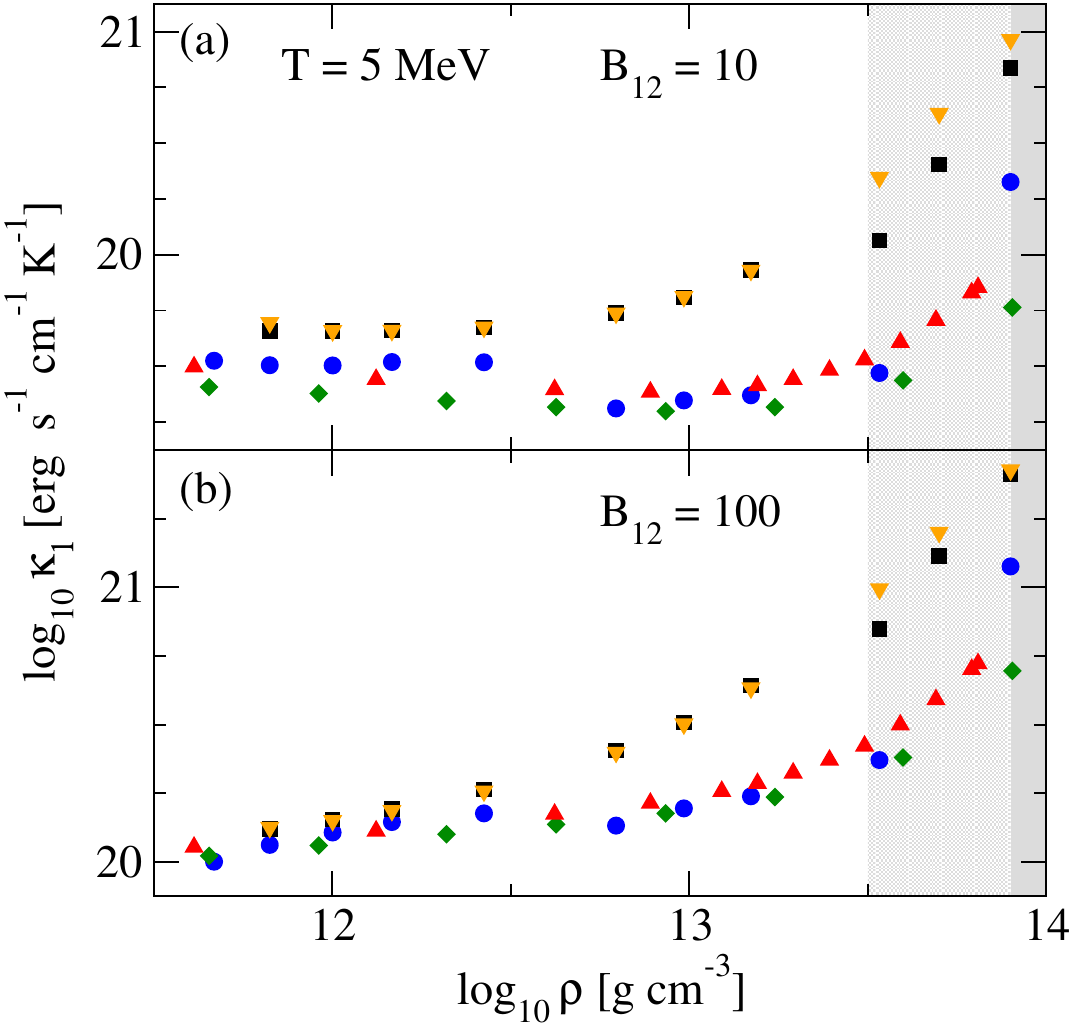}
  \caption{ The same as in  Fig.~\ref{fig:kappa_dens} but for
    $\kappa_1$ and for two magnetic fields $B_{12}=10$ and
    $B_{12}=100$. The shaded regions are the same as in Fig.~\ref{fig:radii}.
  }
\label{fig:kappa1_dens}
\end{center}
\end{figure}
%-------------------------------------------------------
We start with the results concerning the density dependence of the
scalar thermal conductivity, $\kappa$, at a fixed
temperature. Figure~\ref{fig:kappa_dens} shows $\kappa$ as a function
of density for $T = 5$~MeV. Despite the non-monotonic behavior of the
relaxation time with density, the increasing density of states near
the Fermi surface leads to a growth of the thermal conductivity with
matter density, as also suggested by the low-temperature formula
\eqref{eq:kappa_tilde_deg} for $\kappa$.  Since $Z$ varies little
across the inner crust, remaining fixed at a (semi)magic number—the
density dependence of $\kappa$ is governed primarily by the variation
of $A$, which influences the thermal conductivity through the nuclear
form factor. As shown in Fig.~\ref{fig:kappa_dens}, the discrepancies
among different compositions grow with increasing density, reaching
factors 3-4 at $\log_{10} \rho \geq 13$.

The temperature dependence of the thermal conductivity for the D1M*
model at fixed densities is shown in Fig.~\ref{fig:kappa_temp}. In the
range $0 \le T \le 10$~MeV, the thermal conductivity increases with
temperature, consistent with the low-$T$ asymptotic
formula~\eqref{eq:kappa_tilde_deg}. At low densities, where the
relaxation time depends weakly on temperature, $\kappa$ increases
almost linearly with $T$. In the high-density regime, where the effect
of the form factor is more pronounced, we find approximate scaling
$\kappa \propto T^{0.5}$. We note that the asymptotic
formula~\eqref{eq:kappa_tilde_deg} is not exact at $T \simeq 10$~MeV
and moderate densities $\log_{10} \rho \leq 13$, where electrons are
semi-degenerate. In this case, $\kappa$ grows faster than linearly due
to the additional increase in electron energy, which expands the phase
space for thermal conduction.

\subsection{Transverse and Hall conductivities}

Now we consider strong magnetic fields, where the transverse and Hall
components of the thermal conductivity become
relevant. Figure~\ref{fig:kappa0_dens} shows the density dependence of
the $\kappa_0$ component for two values of the magnetic field
($B_{12} = 10, 100$) for selected compositions at $T =
5$~MeV. Figure~\ref{fig:kappa1_dens} presents the same for the
$\kappa_1$ component.

For $B_{12} = 10$, we have $\omega_c \tau \ll 1$, corresponding to
essentially isotropic conduction. In this case, $\kappa_0$ is almost
identical to the scalar conductivity $\kappa$, as seen by comparing
Figs.~\ref{fig:kappa_dens} and \ref{fig:kappa0_dens}, while $\kappa_1$
remains much smaller.  For $B_{12} = 100$, anisotropy is already
significant, as also indicated by Fig.~\ref{fig:tau_dens}. Here, the
transverse and Hall conductivities, $\kappa_0$ and $\kappa_1$, are of
comparable magnitude.
%------------------
\begin{figure}[t] 
\begin{center}
\includegraphics[width=0.95\linewidth, keepaspectratio]{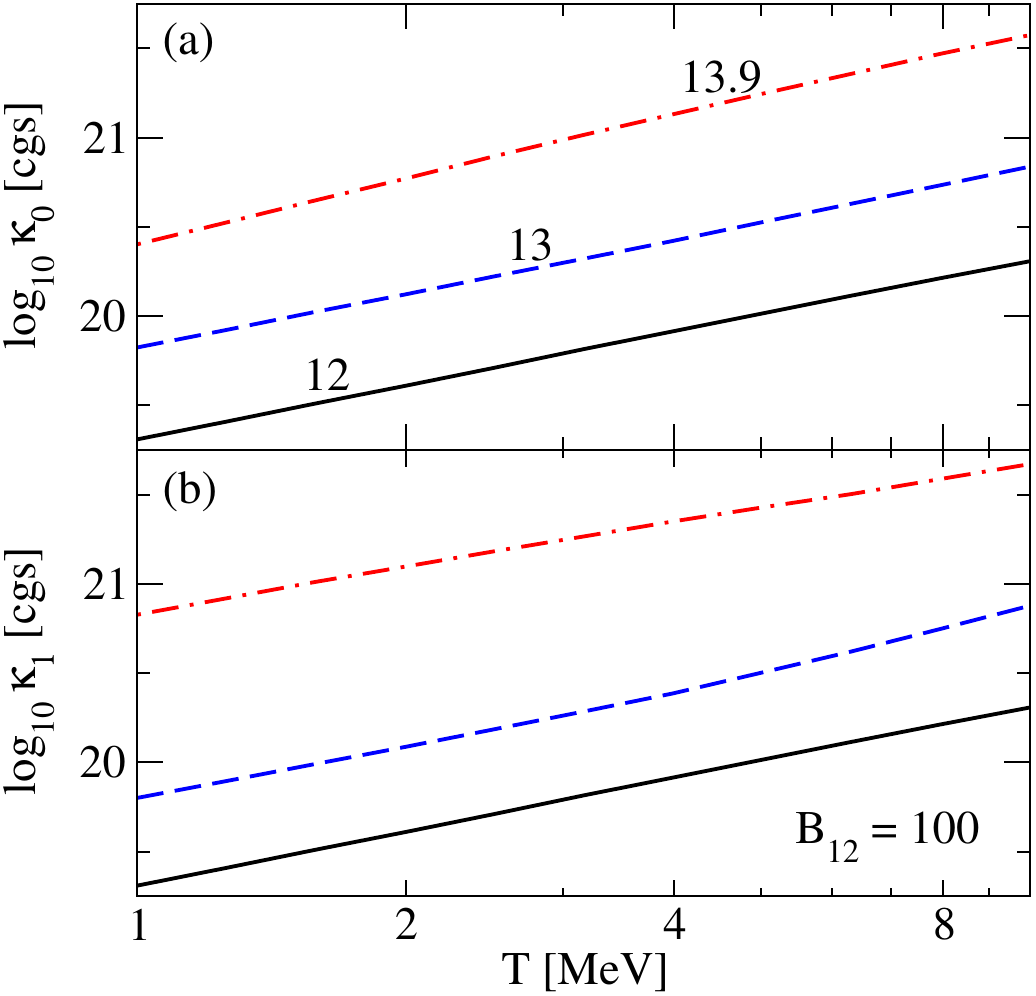}
\caption{ Temperature dependence of $\kappa_0$ and $\kappa_1$ at fixed
  magnetic field value $B_{12}=100$.  Curves correspond to several
  fixed densities, distinguished in the plot by the logarithm of their
  values, for the D1M* composition. 
  }
\label{fig:kappa01_temp}
\end{center}
\end{figure}
%-------------------
\begin{figure}[hbt] 
\begin{center}
\includegraphics[width=0.95\linewidth, keepaspectratio]{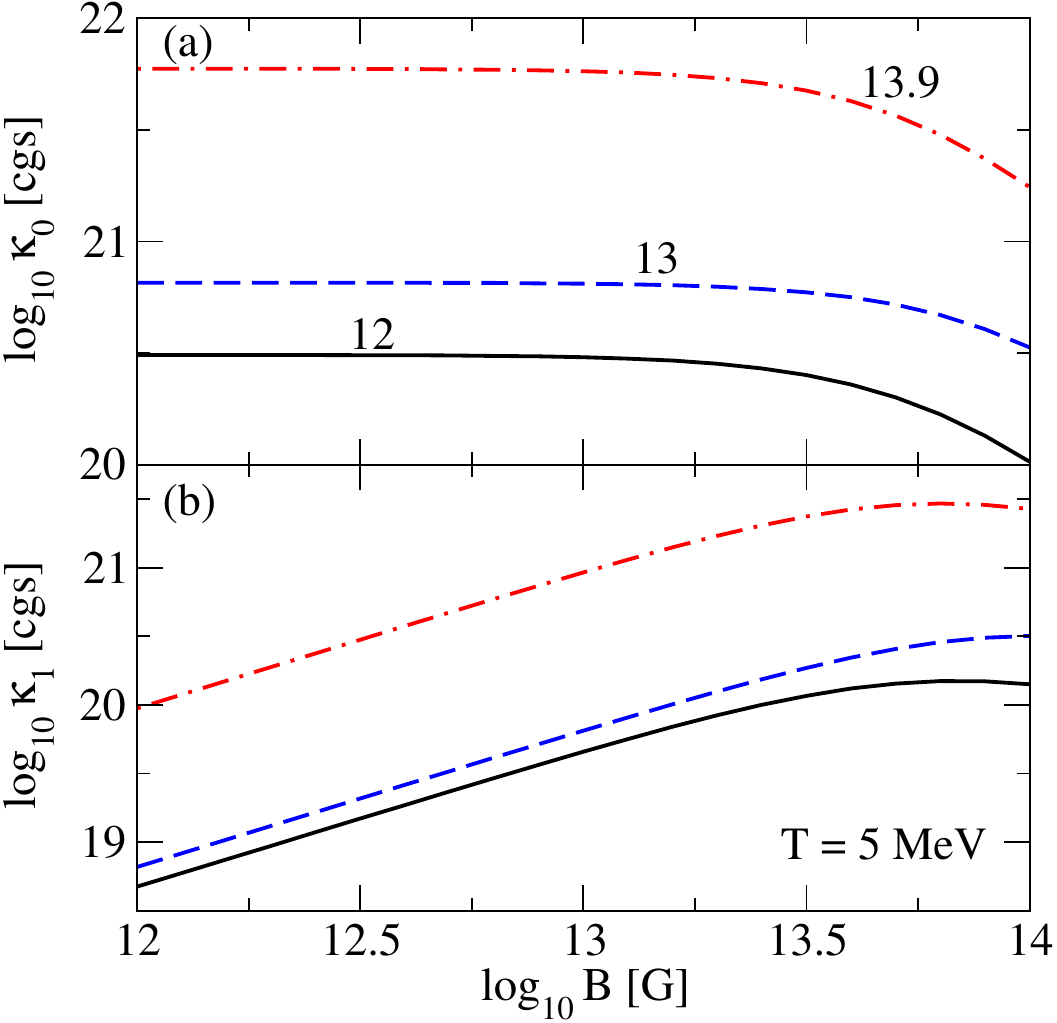}
\caption{ Magnetic field dependence of $\kappa_0$ and $\kappa_1$ at
  fixed temperature $T=5$~MeV. Curves correspond to several
  fixed densities, distinguished in the plot by the logarithm of their
  values, for the D1M* composition. }.
  
\label{fig:kappa_b}
\end{center}
\end{figure}
%----------------

The temperature dependence of $\kappa_0$ and $\kappa_1$ for the D1M*
model is shown in Fig.~\ref{fig:kappa01_temp}, with $B_{12} = 100$ --
a value of the $B$--field for which $\omega_c \tau \sim 1$. Both
components grow nearly linearly with temperature, as implied by
Eq.~\eqref{eq:kappa_tilde_deg}, since the numerators and denominators
in the fractions containing $\tau$ effectively cancel.

Figure~\ref{fig:kappa_b} shows the magnetic field dependence of these
components at fixed temperature and density. According to
Eq.~\eqref{eq:kappa_isotrop}, in the low-field regime $\kappa_1$ is
proportional to the magnetic field, whereas $\kappa_0$ is essentially
independent of it; these trends are visible in
Fig.~\ref{fig:kappa_b}. Anisotropy becomes important for
$B_{12} \gtrsim 30$, where $\kappa_0$ begins to decrease, and
$\kappa_1$ approaches its maximum. Overall, the effect of the magnetic
field on thermal conductivity in the inner crust up to
$B_{12} \simeq 100$ is less pronounced than in the outer crust, which
becomes fully anisotropic for
$B_{12} \gtrsim 10$~\cite{Harutyunyan2016,Harutyunyan2024-2}.

\subsection{Thermopower}

\begin{figure}[hbt]
 \begin{center}
\includegraphics[width=0.95\linewidth,keepaspectratio]{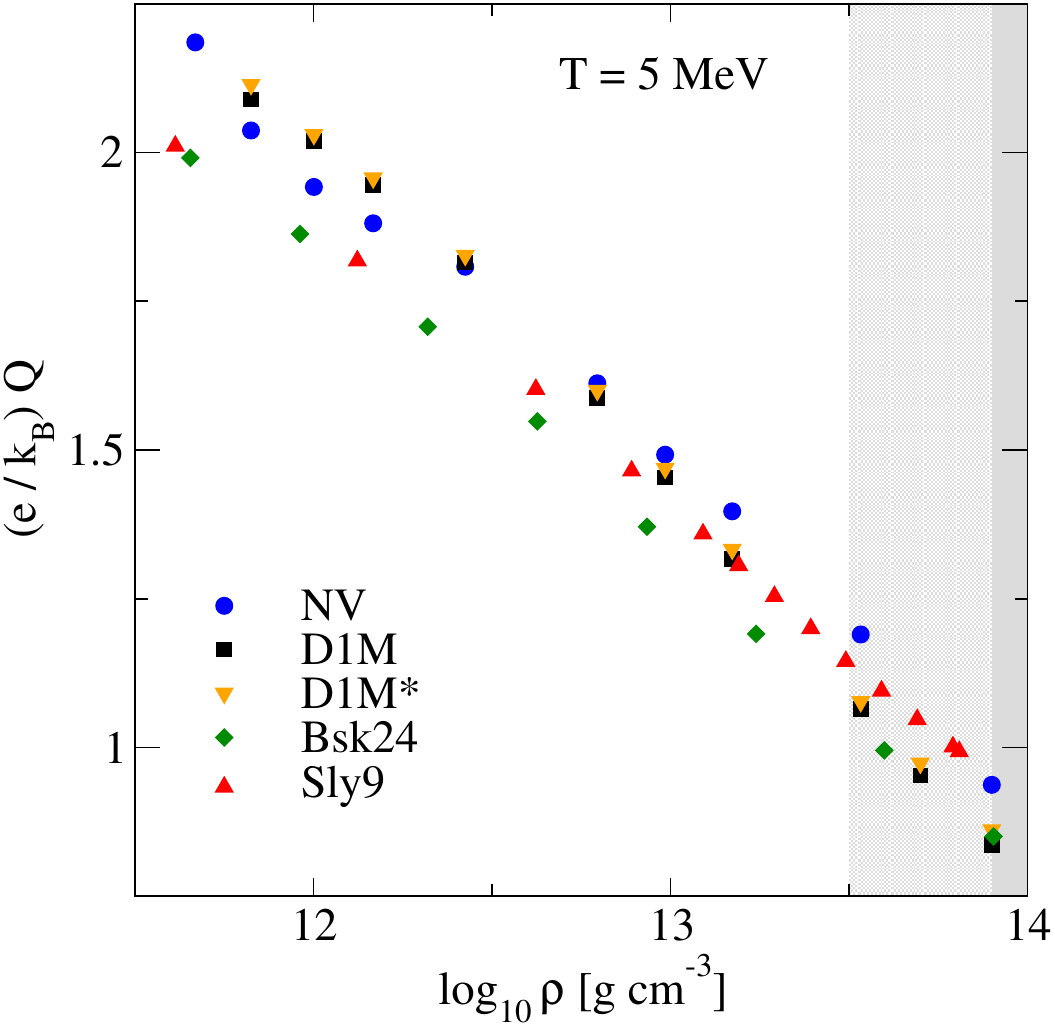}
\caption{ Dependence of thermopower $Q$ on density for five
  compositions. The temperature is fixed at $T=5$~MeV.  The shaded
  regions are the same as in Fig.~\ref{fig:radii}.
} \label{fig:Q_dens}
  \end{center}
\end{figure}

\begin{figure}[hbt]
 \begin{center}
    \includegraphics[width=0.95\linewidth,keepaspectratio]{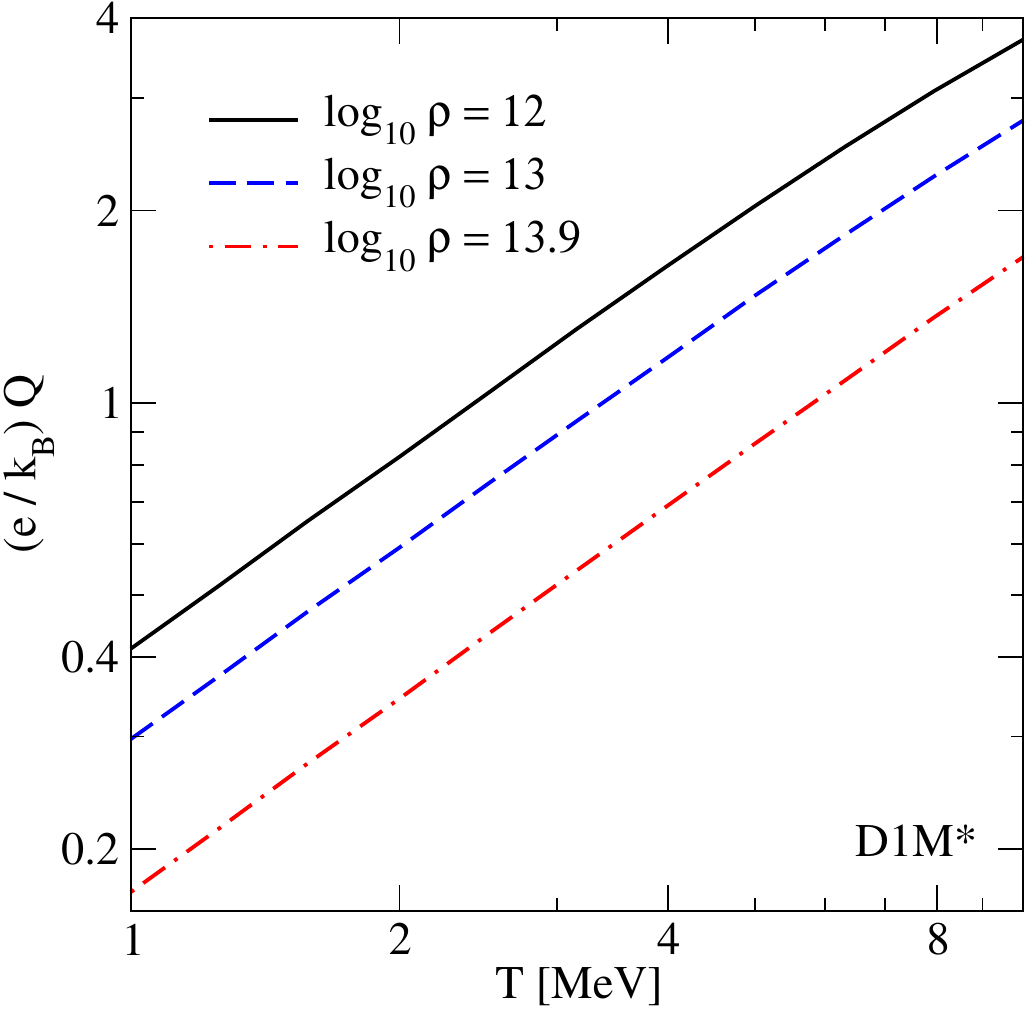} 
    \caption{Dependence of thermopower $Q$ on 
      temperature for various densities, as indicated in the plot,
      for the D1M* composition. 
          } \label{fig:Q_temp}
  \end{center}
\end{figure}

The dependence of the longitudinal thermopower, $Q$, on density and
temperature is shown in Figs.~\ref{fig:Q_dens} and \ref{fig:Q_temp},
respectively. In essence, $Q$ is a thermodynamic quantity and is
practically independent of the microscopic relaxation time. Indeed,
the $\tau$-dependence in $Q$ arises from both $\alpha$ and $\sigma$,
which almost entirely cancel each other, as seen from
Eq.~\eqref{eq:Q_comp}. Consequently, the behavior of $Q$ differs
quantitatively from that of the conductivities. In contrast to the
thermal conductivity, the thermopower decreases with density. This
reversed density dependence is evident from the low-temperature
formula~\eqref{eq:Q_deg1}, which implies the scaling
$Q \propto T n_i^{-1/3}$. This scaling also explains the weak
composition dependence of $Q$ up to the highest densities in the
crust, clearly seen in Fig.~\ref{fig:Q_dens}. As a function of
temperature, thermopower exhibits an almost linear scaling, consistent
with the low-$T$ asymptotics, as shown in Fig.~\ref{fig:Q_temp}.
% -------------------------------------------------------
% -------------------------------------------------------
% -------------------------------------------------------

\begin{figure}[b] \begin{center}
\includegraphics[width=0.95\linewidth,keepaspectratio]{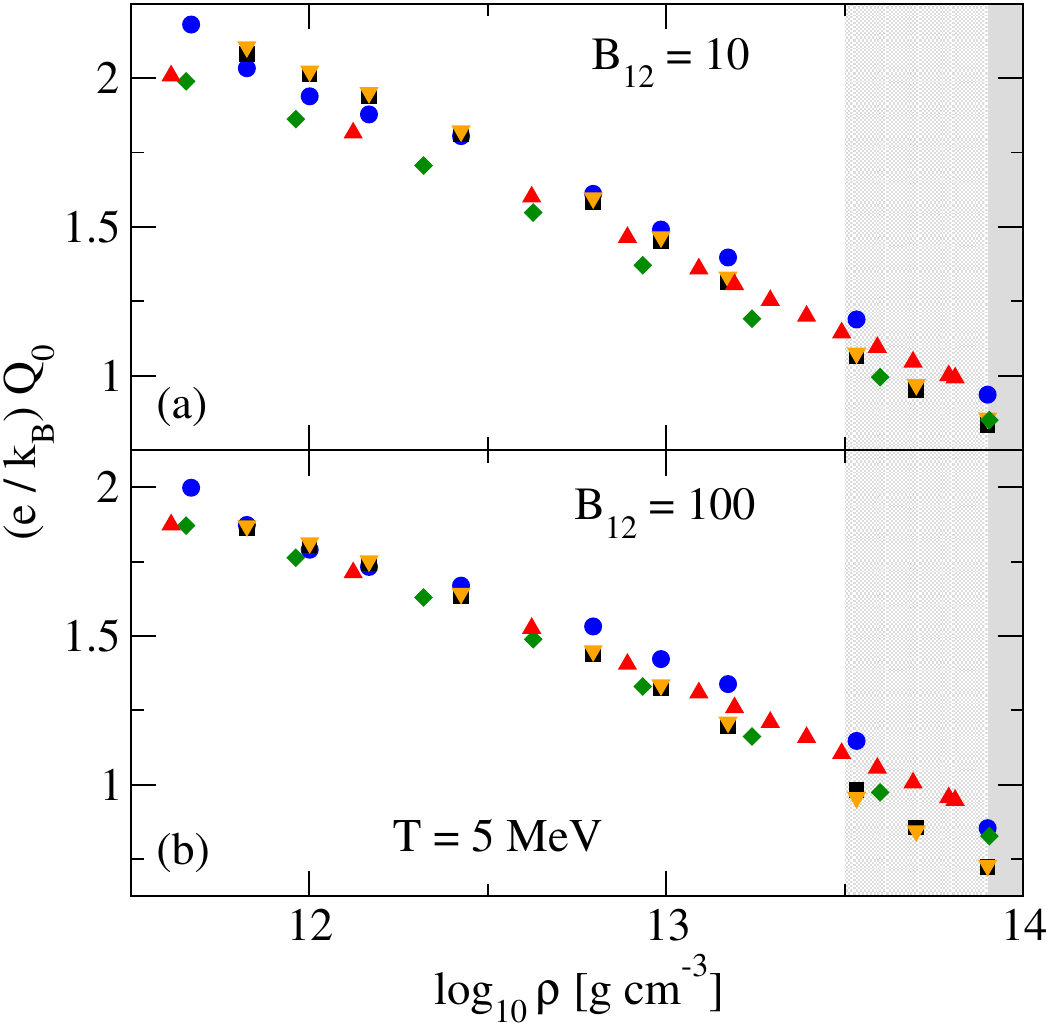}
\caption{ Dependence of the $Q_0$ component of thermopower
  on density for five compositions. The values of the temperature and
  the magnetic field are indicated in the plot. The shaded regions are the
same as in Fig.~\ref{fig:radii}.
} 
\label{fig:Q0_dens} \end{center}
\end{figure}

\begin{figure}[hbt]
  \begin{center}
    \includegraphics[width=0.95\linewidth,keepaspectratio]{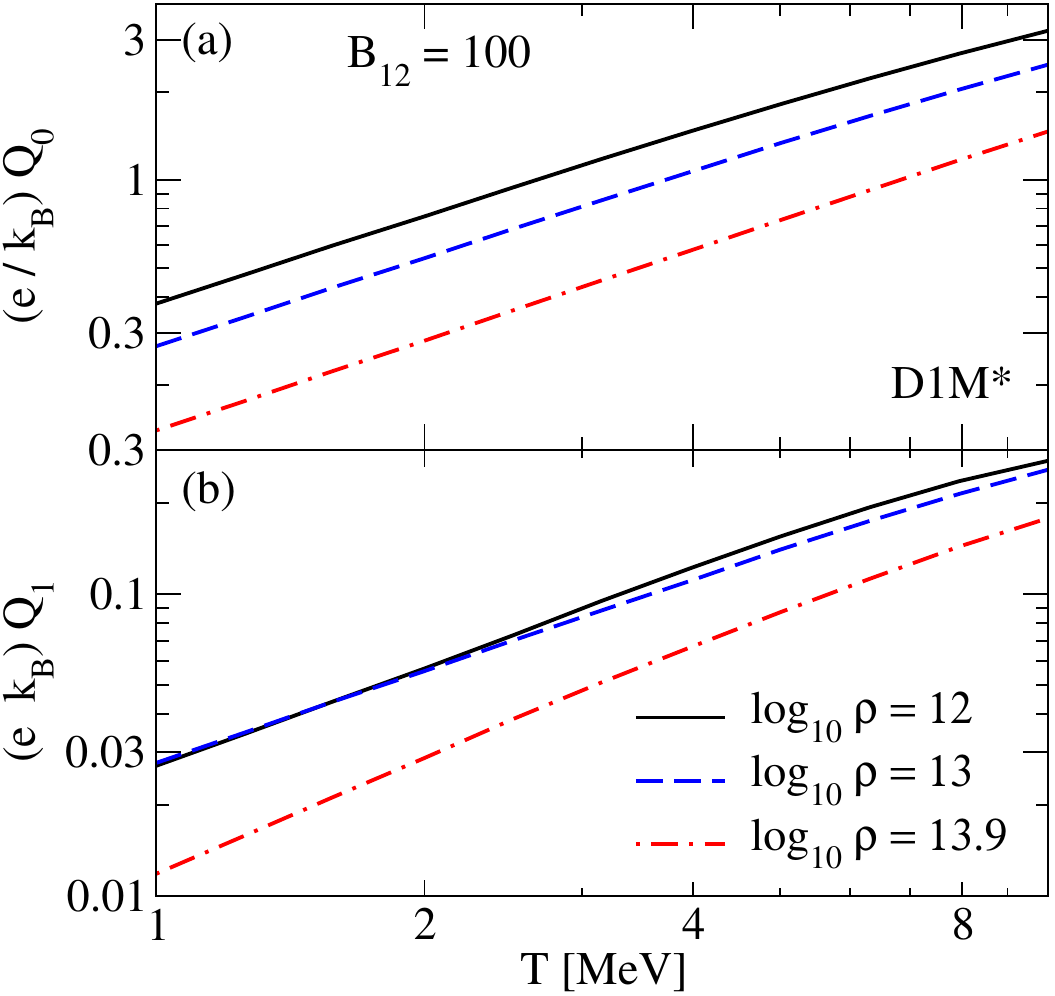} 
    \caption{ Dependence of $Q_0$ and $Q_1$ components of thermopower
      on temperature for various values of density for model D1M*.
    } \label{fig:Q01_temp}
  \end{center}
\end{figure}

\begin{figure}[hbt]
  \begin{center}
    \includegraphics[width=0.95\linewidth,keepaspectratio]{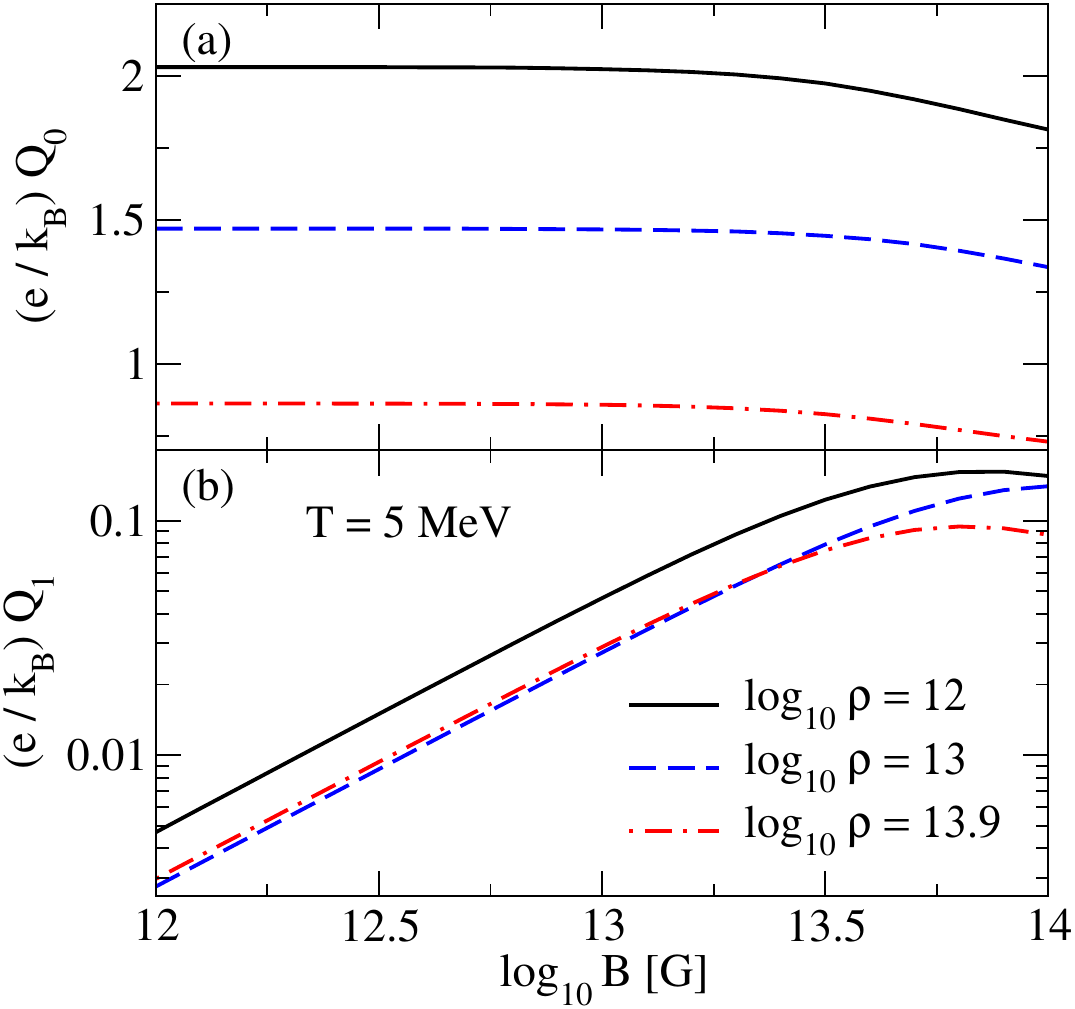}
    \caption{ Dependence of $Q_0$ and $Q_1$ components of thermopower
      on the magnetic field for various values of the density for
      D1M*.  
      } \label{fig:Q01_b}
  \end{center}
\end{figure}

The transverse component, $Q_0$, does not differ significantly from
the longitudinal component $Q$, as implied by the limiting
formula~\eqref{eq:Q_deg2}. Its density and temperature dependence is
nearly identical to that of $Q$, and its magnetic field dependence is
very weak throughout the regime of interest (Figs.~\ref{fig:Q0_dens},
\ref{fig:Q01_temp}, and \ref{fig:Q01_b}).
Specifically, Eqs. \eqref{eq:Q_deg1} and \eqref{eq:Q_deg2} give
$Q_0\to Q$ for $\omega_c\tau\ll1$ and $Q_0\to3Q/4$ for
$\omega_c\tau\gg1$, respectively.

\begin{figure}[b] \begin{center}
\includegraphics[width=0.95\linewidth,keepaspectratio]{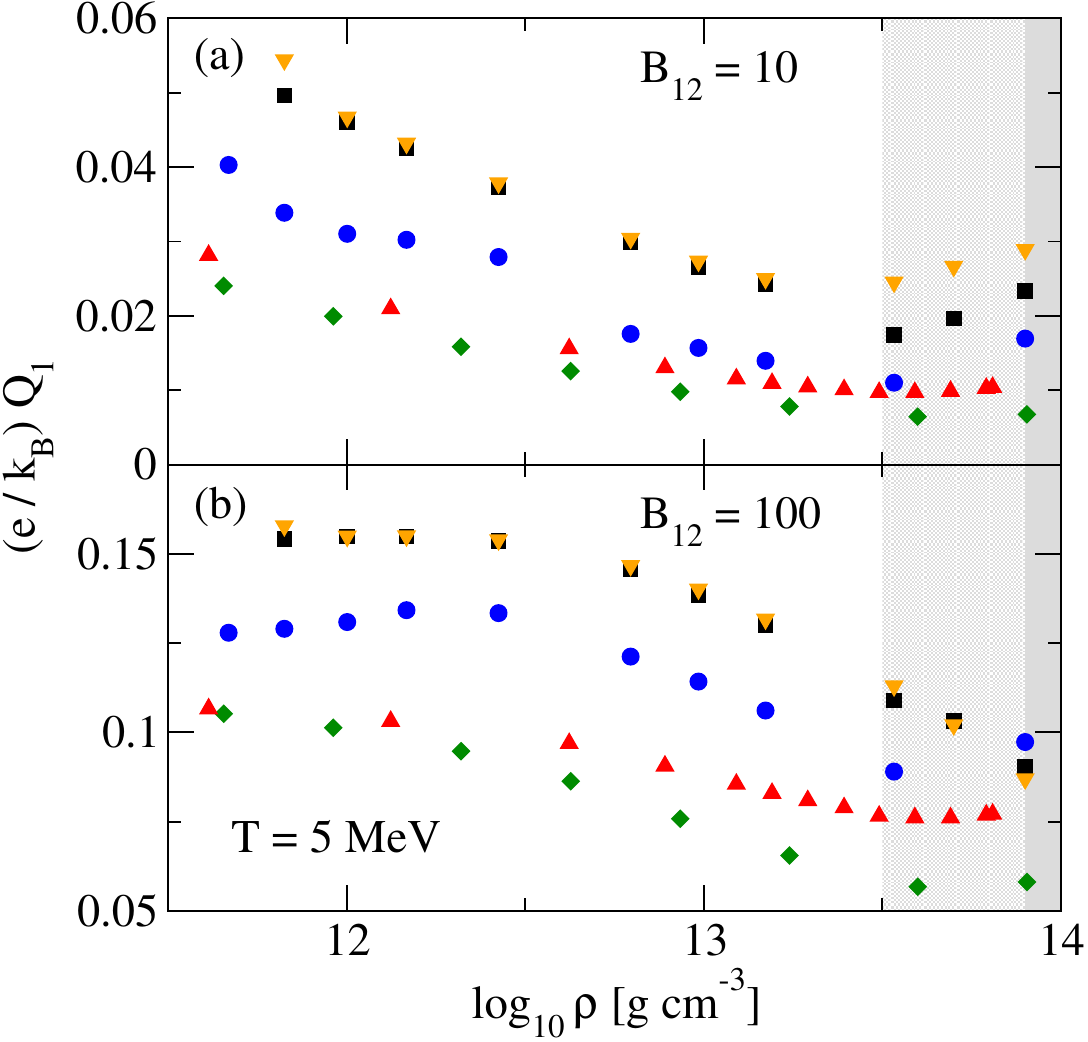}
\caption{Dependence of the Hall thermopower $Q_1$ on density for five
  compositions. The values of the temperature and the magnetic field
  are indicated in the plot.  The shaded regions are the
    same as in Fig.~\ref{fig:radii}.}
\label{fig:Q1_dens} \end{center}
\end{figure}

The Hall component of thermopower, $Q_1$, is shown as a function of density in Fig.~\ref{fig:Q1_dens}. As seen, $Q_1$ differs significantly among compositions: D1M and D1M* predict higher values, Bsk24 and Sly9 predict smaller values, and NV is intermediate. This scatter reflects the nontrivial dependence of $Q_1$ on the relaxation timescale, which does not cancel in contrast to $Q_0$, as seen from Eq.~\eqref{eq:Q_deg3}.
% -------------------------------------------------------
% -------------------------------------------------------
% -------------------------------------------------------

The temperature dependence of $Q_1$ at fixed densities and magnetic
field is shown in Fig.~\ref{fig:Q01_temp}. Although $Q_1$  increases
with temperature more rapidly than $Q_0$, it remains smaller in
magnitude. Figure~\ref{fig:Q01_b} presents the magnetic-field
dependence of $Q_1$  for three representative densities at fixed
temperature. The behavior of $Q_1$ closely mirrors that of
$\kappa_1$: it grows with increasing field strength up to $B_{12}
\simeq 30$,
where it reaches a maximum.

% -------------------------------------------------------
% -------------------------------------------------------
% -------------------------------------------------------

\subsection{Magnetic field evolution timescales}
\label{sec:Ohmic_Hall}

%------------------------------------------------------
\begin{figure*}[hbt] \begin{center}
\includegraphics[width=0.48\linewidth,keepaspectratio]{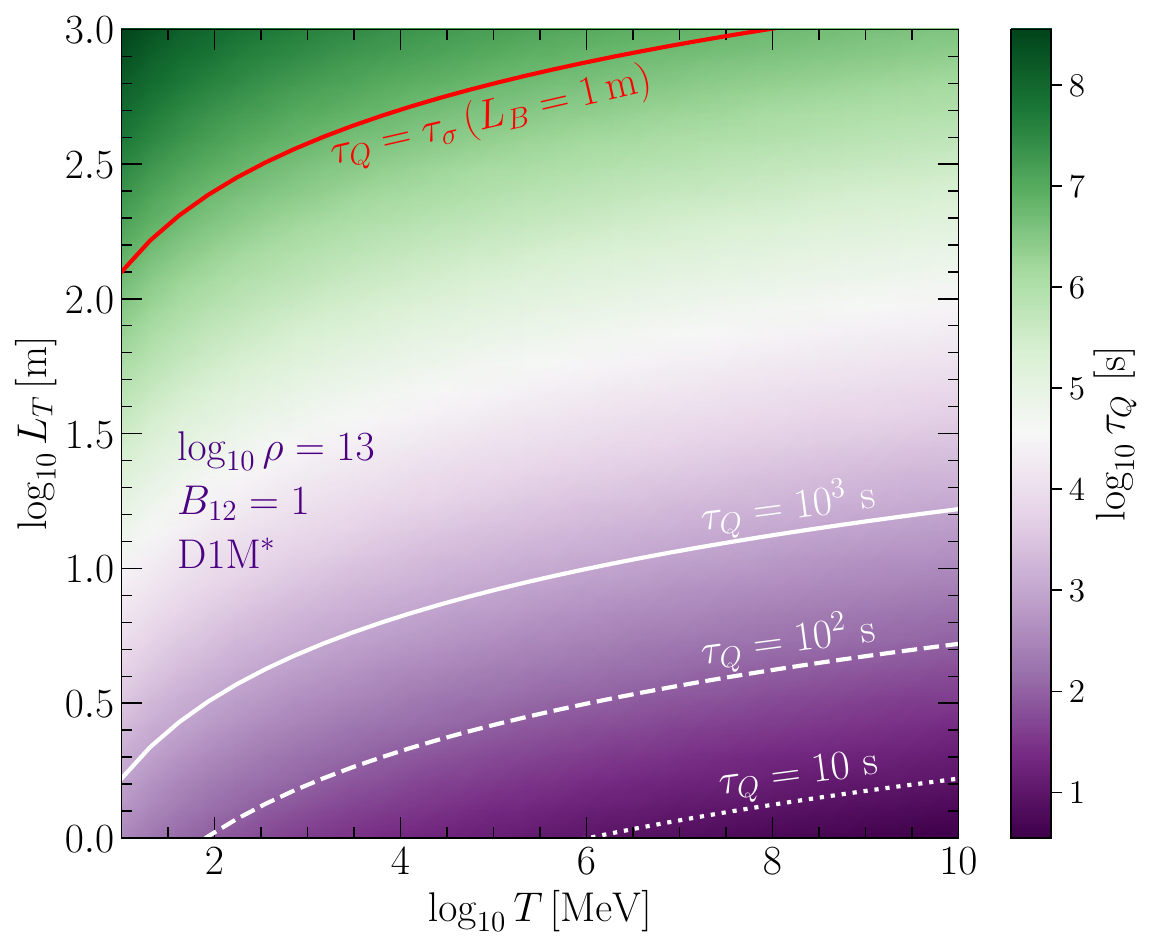} 
\hspace{0.3cm}
\includegraphics[width=0.48\linewidth,keepaspectratio]{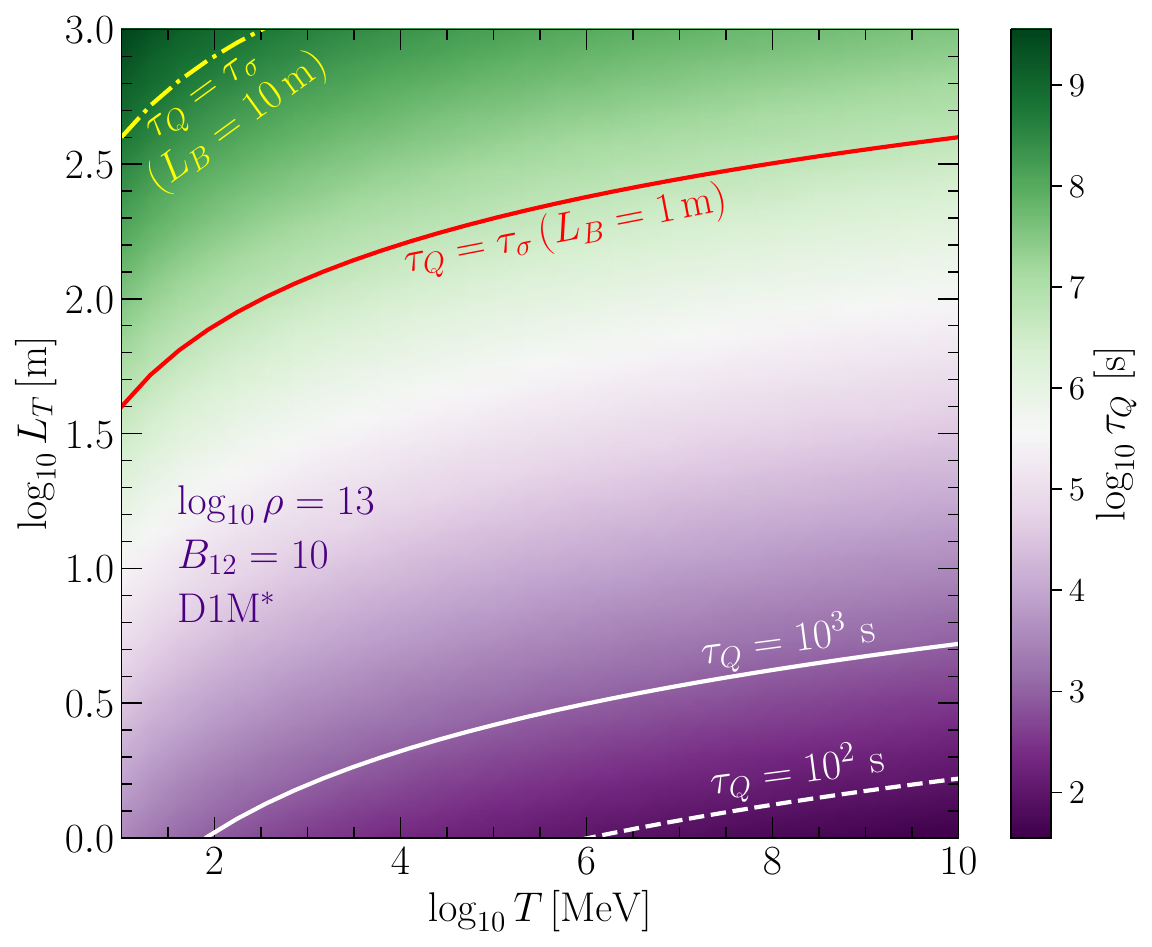} 
\caption{ Magnetic field evolution time scale $\tau_Q$ due to
  thermoelectric effects as a function of temperature and the thermal
  gradient length scale $L_T$ for model D1M*. Magnetic field is fixed
  at $B_{12}=1$ (left figure) and $B_{12}=10$ (right figure). The
  density is fixed at $\rho=10^{13}$\,g\,cm$^{-3}$.  The red and
  yellow curves represent the lines where $\tau_Q=\tau_\sigma$ at
  $L_B=1$\,m and $L_B=10$\,m, respectively. The white curves are the
  lines where $\tau_Q$ reaches $10^3$\,s (solid), $10^2$\,s
  (dashed) and 10\,s (dotted).
}
\label{fig:timescales}
\end{center}
\end{figure*}
%--------------------------------------------------------

We now present simple estimates of the characteristic timescales governing magnetic-field evolution, based on our numerical results for the thermopower and on the electrical conductivity calculated in Ref.~\cite{Harutyunyan2024-1}. The evolution of low-frequency magnetic
fields in neutron stars is governed by Maxwell's equations:
%-----------------------------------------
\bea\label{eq:Maxwell_eqs}
\bm{\nabla} \times \vecE = - \frac{\partial \vecB}{\partial t}, \qquad
\bm{\nabla} \times \vecB = 4\pi \vecj,
\eea
%-----------------------------------------
where we assume the magnetic permeability of matter is unity and neglect the displacement current. Substituting the electric field from
Eq.~\eqref{eq:currents_reversed}, $\vecE = \hat{\varrho} \vecj -
\hat{Q} \bm{\nabla} T - \bm{\nabla} \mu / e$, we obtain the induction equation
including thermoelectric effects:
%-----------------------------------------
\bea\label{eq:mag_field}
\frac{\partial \vecB}{\partial t} &=& - \bm{\nabla} \times \left( \hat{\varrho} \vecj - \hat{Q} \bm{\nabla} T - \frac{1}{e} \bm{\nabla} \mu \right) \nonumber\\
&=& - \frac{1}{4\pi} \bm{\nabla} \times \left( \hat{\varrho} \bm{\nabla} \times \vecB \right) + \bm{\nabla} \times (\hat{Q} \bm{\nabla} T),
\eea
%-----------------------------------------
where we used $\bm{\nabla} \times \bm{\nabla} \mu = 0$.
For moderate fields, $B_{12} \le 10$, Eq.~\eqref{eq:mag_field}
can be taken in the isotropic approximation:
%-----------------------------------------
\bea\label{eq:mag_field_is}
\frac{\partial \vecB}{\partial t} = - \frac{1}{4\pi} \bm{\nabla} \times \left( \frac{\bm{\nabla} \times \vecB}{\sigma} \right) + \bm{\nabla} \times (Q \bm{\nabla} T).
\eea
%-----------------------------------------
To estimate the relevance of the dissipation for the BNS mergers and
other transients, we note that the variations of the conductivity with
the density and temperature are weak and, therefore, we approximate it
by a characteristic constant value
$\sigma = 4\cdot10^{22}$\,s$^{-1}$.  Then, using
$\bm{\nabla} \times (\bm{\nabla} \times \vecB) = \bm{\nabla}
(\bm{\nabla} \cdot \vecB) - \Delta \vecB$ and
$\bm{\nabla} \cdot \vecB = 0$, we obtain:
%-----------------------------------------
\bea\label{eq:mag_field_is1}
\frac{\partial \vecB}{\partial t} = \frac{1}{4\pi \sigma} \Delta \vecB + \bm{\nabla} Q \times \bm{\nabla} T.
\eea
%-----------------------------------------
An estimate of the magnetic field variation timescale, $\tau_B$,
follows from approximating
$|\partial \vecB / \partial t|\simeq B/\tau_B$,
$|\Delta \vecB| \simeq B / L_B^2$, $|\bm{\nabla} Q| \simeq Q / L_T$,
and $|\bm{\nabla} T| \simeq T / L_T$, where $L_B$ and $L_T$ are the
length scales over which the magnetic field and the temperature vary,
respectively. With these approximations we obtain:
%-----------------------------------------
\bea\label{eq:mag_field_is2}
\frac{B}{\tau_B} = \bigg| \frac{B}{4\pi \sigma L_B^2} \pm \frac{Q T}{L_T^2} \bigg|,
\eea
%-----------------------------------------
or alternatively
%-----------------------------------------
\bea\label{eq:decay_time}
\tau_B^{-1} = \big| \tau_\sigma^{-1} \pm \tau_Q^{-1} \big|, \quad
\tau_\sigma = 4\pi \sigma L_B^2, \quad
\tau_Q = \frac{B L_T^2}{Q T},
\eea
%-----------------------------------------
with $\tau_B$ being the field decay timescale, which includes thermoelectric effects,  $\tau_\sigma$ is the Ohmic dissipation timescale, and $\tau_Q$ is a
timescale associated with thermoelectric effects. 
Clearly, the characteristic timescale of variations of the neutron star crustal field is governed by the smaller of these two.

Figure~\ref{fig:timescales} shows the timescale $\tau_Q$ as a function
of temperature $T$ and the length scale $L_T$, for two values of the
magnetic field, $B_{12}=1$ (left panel) and $B_{12}=10$ (right panel),
assuming the composition of model D1M*. The range of $T$ and $L_T$ in
Fig.~\ref{fig:timescales} has been chosen to match the conditions
relevant for binary neutron star mergers. We display results for a
single density, since the dependence of $\tau_Q$ on density is found
to be weak. To gauge the relative importance of the two channels of
dissipation, we show lines on which the Ohmic dissipation timescale
$\tau_\sigma$ becomes equal to $\tau_Q$ for fixed values $L_B=1$\,m
and $L_B=$~10\,m. For larger values of $L_B$, $\tau_\sigma$ exceeds
$\tau_Q$ across the entire range of parameters $T$ and $L_T$
considered. We do not examine scales shorter than a meter, as these
fall well below the resolution of typical numerical
simulations. However, at such small scales, the timescale could become
even shorter and potentially more relevant for the dynamics.

With the adopted length scales, it is seen that the high-temperature,
small-$L_T$ region of the considered parameter range exhibits values
of $\tau_Q$ that are comparable to the dynamical timescales of
long-lived BNS merger remnants, estimated to lie in the range
10–100~s. This region is larger for smaller magnetic fields, i.e.,
$B_{12} = 1$ in the figure. In this regime, Ohmic dissipation plays a
negligible role. However, in the domain corresponding to a large $L_T$
(more appropriate for the resolution of current simulations), $\tau_Q$
increases, and beyond the line $\tau_Q = \tau_\sigma$, the thermopower
becomes unimportant and the evolution is driven by electrical
conductivity.

\section{Summary}
\label{sec:summary}

In this work, we calculated the thermal conductivity and thermopower
at densities corresponding to the inner crusts of neutron stars, in
the regime where ions are in a liquid state. In this regime, electrons
are the main carriers of charge and heat, and the dominant channel of
dissipation is their scattering off correlated nuclei via screened
electromagnetic interactions. The Boltzmann kinetic equation was
solved in the relaxation time approximation in the simultaneous
presence of electromagnetic fields and thermal gradients. We analyzed
the general tensor structure of the transport coefficients, including
the effects of ion-ion correlations and finite nuclear
size. Additionally, we computed the exact rate of electron scattering
on the anomalous magnetic moment of free neutrons at finite
temperatures and found that its contribution to electron transport is
strongly suppressed.

For numerical calculations, five different nuclear compositions of the
inner crust were considered. The nuclear form factor was found to be
the dominant factor governing both the magnitude of the thermal
conductivity and its scatter among different compositions at high
densities. Thermal conduction becomes anisotropic for magnetic fields
$B_{12} \gtrsim 30$, with the transverse and Hall components
suppressed relative to the longitudinal component. In contrast,
thermopower exhibits more universal behavior, being largely
independent of the relaxation time.

Using these results for conductivity and thermopower, we estimated the
magnetic field evolution timescales in the neutron star inner crust,
including thermoelectric effects. Our analysis shows that
thermoelectric effects can become important in the magnetic field
evolution of hot post-merger remnants and proto-neutron stars at
moderate values of magnetic fields, $B_{12}\sim 1$, if temperature
exhibits gradients at sufficiently small scales of the order of a
meter. At low temperatures and larger magnetic fields, the role of
thermoelectric effects in the dynamical evolution of these objects
becomes unimportant.

The quantitative scope of these conclusions is set by the ionic model.
We used static correlations of a point-ion OCP, incorporated finite
nuclear size only through the single-ion scattering form factor, and
kept zero-temperature single-species compositions fixed as the
temperature was varied. The small thermal velocity of heavy nuclei
supports the static approximation as the leading heavy-ion limit, but
a controlled correction for ionic dynamics requires
$S(q,\omega)$. Likewise, the high-temperature end of our results,
especially $T\le 10$~MeV, should be interpreted as transport for a
prescribed composition rather than as a self-consistent hot-crust
model. Finite-temperature multi-component composition, modification of
ion correlations by screening and finite charge extent, cluster
dissolution, and pasta-like structures can all become important in the deep
crust or in accreting and merger environments. Extending the present
framework to partial static/dynamical structure factors of a
multi-component plasma is therefore a natural next step; the present
results provide the corresponding single-component reference study.

\section*{Acknowledgments}

The authors acknowledge support from the collaborative research grant
No. 24RL-1C010 provided by the Higher Education and Science Committee
(HESC) of the Republic of Armenia through the “Remote Laboratory”
program. A.~S. also acknowledges support from the Deutsche
Forschungsgemeinschaft Grant No. SE 1836/6-1 and the Polish National
Science Centre (NCN) Grant No. 2023\-/51/B/ST9/02798.

% -----------------------------------------------------------------------------
% EPJ A currently asks for relevant Statements and Declarations in the manuscript.
% Complete/uncomment the items that apply before submission.  The wording below is
% intentionally left as a placeholder rather than assuming declarations for the authors.
%-----------------------------------------------------------------------------
 \section*{Statements and Declarations}
 \subsection*{Competing Interests}
The authors declare no competing interests.

 \subsection*{Data availability}
 The data and results supporting the findings of this study are
 available from the corresponding author upon reasonable request.
% -----------------------------------------------------------------------------

\appendix

\section{Electron-neutron scattering matrix element}
\label{app:matrix_en}

\begin{figure}[t]
\centering
\includegraphics[width=\columnwidth,keepaspectratio]{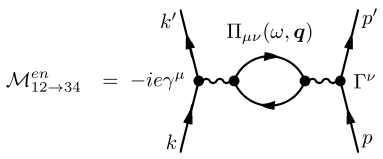}
\caption{Diagram describing electron-neutron scattering via exchange of
a virtual plasmon. The plasmon self-energy is given by the polarization
tensor $\Pi_{\mu\nu}(\omega,\vecq)$, shown by the closed loop.}
\label{fig:diagram}
\end{figure}

In this Appendix, we compute the leading-order matrix element for
electron--neutron scattering via virtual plasmon exchange, where the
neutron couples to the plasmon only through its anomalous magnetic
moment.  The Lagrangian of this interaction is
%-------------------------------------
\bea\label{Aeq:1}
\mathcal{L}_{\rm int}
=
-\frac{e \kappa_n}{4 m_n}
\bar{\psi}_n \sigma^{\mu \nu} \psi_n F_{\mu \nu},\quad 
\sigma^{\mu\nu}
=
\frac{i}{2}
\left(\gamma^{\mu}\gamma^{\nu}
-\gamma^{\nu}\gamma^{\mu}\right),\nonumber\\
\eea
%-------------------------------------
where $\kappa_n \approx-1.91$ is the neutron anomalous magnetic moment
in nuclear magnetons, $m_n$ is the neutron mass, and $F_{\mu \nu}$ is
the electromagnetic field tensor. The effective vertex for
electron-neutron scattering follows from Eq.~\eqref{Aeq:1},
%---------------------
\bea\label{Aeq:2}
\Gamma^\mu
=
i \frac{e \kappa_n}{2 m_n}\sigma^{\mu \nu} q_\nu ,
\eea
%---------------------
where $q^\nu$ is the plasmon momentum entering the vertex, \ie, the
momentum transfer in the scattering process.  In a thermal medium, this
process is mediated by a screened interaction corresponding to plasmon
exchange, as illustrated in Fig.~\ref{fig:diagram}; see
Ref.~\cite{Harutyunyan2016} and references therein for a more detailed
discussion of the HTL approximation.  We split the matrix element into
longitudinal and transverse parts,
%-------------------------
\bea\label{eq:matrix_en}
{\cal M}^{en}_{12\rightarrow34}
&=&
\frac{ie^2\kappa_n}{2m_n}
\left({\cal M}_L-{\cal M}_T\right),
\nonumber\\
{\cal M}_L
&=&
\frac{J_e^0J_n^0}{\vecq^{\,2}+\Pi_L},
\nonumber\\
{\cal M}_T
&=&
\frac{J_{e\perp}^iJ_{n\perp}^i}
{\vecq^{\,2}-\omega^2+\Pi_T},
\eea
%-------------------------
where the electron and neutron currents are defined as
%-------------------------
\bea
J_e^\mu
=
\bar{u}_e(k')\gamma^{\mu}u_e(k),
\quad 
J_n^\nu
=
\bar{u}_n(p')
\left(\sigma^{\nu\lambda}q_{\lambda}\right)u_n(p).
\eea
%-------------------------
Here $k,k^{\prime}$ are the initial and final 4-momenta of the
electron, $p,p^{\prime}$ are the initial and final 4-momenta of the
neutron, $q=(\omega,\vecq)=k-k^{\prime}=p^{\prime}-p$, and
$\bm J_{e\perp}$ and $\bm J_{n\perp}$ are the components of these
currents transverse to $\bm q$.

%-------------------------------------------------------
%-------------------------------------------------------

The screening of the interaction is taken into account through the
longitudinal $\Pi_L(\omega,q)$ and transverse $\Pi_T(\omega,q)$
components of the plasmon polarization tensor. We employ the HTL
results for ultrarelativistic electrons
\cite{Braaten1990,Braaten1992,Harutyunyan2016},
%-----------------------------------------------
\bea\label{eq:pol_deg_ultra_l}
\Pi_L(\omega,q)
&=&
q_D^2
\left[
1-\frac{x}{2}\log\frac{x+1}{x-1}
\right],
\\
\label{eq:pol_deg_ultra_t}
\Pi_T(\omega,q)
&=&
\frac{1}{2}q_D^2
\left[
x^2+
\left(1-x^2\right)
\frac{x}{2}
\log\frac{x+1}{x-1}
\right],
\eea
%-----------------------------------------------
where $x=\omega/|\vecq|$, and $q_D$ is given by Eq.~\eqref{eq:Debye}.
Squaring Eq.~\eqref{eq:matrix_en}, we obtain
%-------------------------
\bea\label{eq:matrix_en1}
|{\cal M}^{en}_{12\rightarrow34}|^2
&=&
\frac{e^4\kappa_n^2}{4m_n^{2}}
\left[
|{\cal M}_L|^2
+|{\cal M}_T|^2
-2{\rm Re}
\left({\cal M}_L{\cal M}_T^*\right)
\right],\nonumber\\
\eea
%-------------------------
with
%-------------------------
\bea\label{eq:matrix_en2}
|{\cal M}_L|^2
&=&
\frac{
J_e^0J_e^{0*}J_n^0J_n^{0*}}
{|\vecq^{\,2}+\Pi_L|^2},
\\
|{\cal M}_T|^2
&=&
\frac{
J_{e\perp}^iJ_{e\perp}^{k*}
J_{n\perp}^iJ_{n\perp}^{k*}}
{|\vecq^{\,2}-\omega^2+\Pi_T|^2},
\\
\label{eq:matrix_en2_lt}
{\cal M}_L{\cal M}_T^*
&=&
\frac{
J_e^0J_{e\perp}^{i*}
J_n^0J_{n\perp}^{i*}}
{(\vecq^{\,2}+\Pi_L)
(\vecq^{\,2}-\omega^2+\Pi_T^{*})}.
\eea
%-------------------------
After averaging over spins, we find
%-------------------------
\bea\label{eq:matrix_en3}
\overline{|\mathcal{M}_L|^2}
&=&
\frac{1}{4}\sum_{\rm spins}
|\mathcal{M}_L|^2
\nonumber\\
&=&
\frac{L^{00}H^{00}}
{64\ep_k\ep_{k'}\ep_p\ep_{p'}
|\vecq^{\,2}+\Pi_L|^2},
\\
\label{eq:matrix_en4}
\overline{|\mathcal{M}_T|^2}
&=&
\frac{1}{4}\sum_{\rm spins}
|\mathcal{M}_T|^2
\nonumber\\
&=&
\frac{L^{ik}_{\perp}H^{ik}_{\perp}}
{64\ep_k\ep_{k'}\ep_p\ep_{p'}
|\vecq^{\,2}-\omega^2+\Pi_T|^2},
\\
\label{eq:matrix_en5}
\overline{{\cal M}_L{\cal M}_T^*}
&=&
\frac{1}{4}\sum_{\rm spins}
{\cal M}_L{\cal M}_T^*
\nonumber\\
&=&
\frac{L^{0i}_{\perp}H^{0i}_{\perp}}
{64\ep_k\ep_{k'}\ep_p\ep_{p'}
(\vecq^{\,2}+\Pi_L)
(\vecq^{\,2}-\omega^2+\Pi_T^{*})}.
\eea
%-------------------------
The leptonic and hadronic tensors are
%-----------------------
\bea\label{eq:lepton}
L_{\mu \nu}
&=&
4\ep_k\ep_{k'}
\sum_{\rm spins}J_{e\mu} J_{e\nu}^{*}
\nonumber\\
&=&
{\rm Tr}
\left[
\left(\slashed{k}^{\prime}+m_e\right)
\gamma_\mu
\left(\slashed{k}+m_e\right)
\gamma_\nu
\right],
\\
\label{eq:hadron}
H^{\mu \nu}
&=&
4\ep_p\ep_{p'}
\sum_{\rm spins}J_n^{\mu}J_n^{\nu*}
\nonumber\\
&=&
{\rm Tr}
\left[
\left(\slashed{p}^{\prime}+m_n\right)
\sigma^{\mu \rho}q_\rho
\left(\slashed{p}+m_n\right)
\sigma^{\nu \sigma}q_\sigma
\right].
\eea
%-----------------------
Using standard trace algebra for Dirac matrices, one obtains
%-----------------------
\bea\label{eq:L_mu_nu}
L^{\mu \nu}
&=&
4\left[
k^\mu k'^\nu+k^\nu k'^\mu
%\right.
%\nonumber\\
%&&
%\left.
-g^{\mu \nu}
\left(k\cdot k^{\prime}-m_e^2\right)
\right],
\\
\label{eq:H_mu_nu}
H^{\mu \nu}
&=&
4 \Big[
(q\cdot p)(q^{\mu}p'^{\nu}+q^{\nu}p'^{\mu})
%\nonumber\\
%&&
+(q\cdot p')(q^{\mu}p^{\nu}+q^{\nu}p^{\mu})
\nonumber\\
&&
-q^2(p'^{\mu}p^{\nu}+p'^{\nu}p^{\mu})
%\nonumber\\
%&&
-2(q\cdot p)(q\cdot p')g^{\mu\nu}
\Big]
\nonumber\\
&&
+4\left(m_n^{2}+p\cdot p'\right)
\left(q^2 g^{\mu \nu}-q^\mu q^\nu\right).
\eea
%-----------------------
The relevant components of the leptonic tensor are
%-------------------------
\bea\label{eq:L_00}
L^{00}
&=&
4\left(\ep_k\ep_{k'}
+\veck\cdot\veck'
+m_e^2\right),
\\
\label{eq:L_0i}
L^{0i}_{\perp}
&=&
4\left(
\ep_k k'^i_{\perp}
+\ep_{k'} k^i_{\perp}
\right),
\\
\label{eq:L_ik}
L^{ik}_{\perp}
&=&
4\Big[
k^i_{\perp}k'^k_{\perp}
+k^k_{\perp}k'^i_{\perp}
\nonumber\\
&&
+\delta^{ik}_{\perp}
\left(
\ep_k\ep_{k'}
-\veck\cdot\veck'
-m_e^2
\right)
\Big],
\eea
%-------------------------
whereas the relevant components of the hadronic tensor are
%-------------------------
\bea\label{eq:H_00}
H^{00}
&=&
8\Big[
\omega\ep_{p'}(q\cdot p)
+\omega\ep_p(q\cdot p')
%\nonumber\\
%&&
-q^2\ep_p\ep_{p'}
-(q\cdot p)(q\cdot p')
\Big]
\nonumber\\
&&
+4\left(m_n^{2}+p\cdot p'\right)
\left(q^2-\omega^2\right),
\\
\label{eq:H_0i_perp}
H^{0i}_\perp
&=&
4 \left[
\omega(q\cdot p)-q^2\ep_p
\right]p'^{i}_\perp
%\nonumber\\
%&&
+4\left[
\omega(q\cdot p')-q^2\ep_{p'}
\right]p^{i}_\perp,
\\
\label{eq:H_ik_perp}
H^{ik}_\perp
&=&
-4q^2
\left(
p'^{i}_\perp p^{k}_\perp
+p'^{k}_\perp p^{i}_\perp
\right)
%\nonumber\\
%&&
+4\Big[
2(q\cdot p)(q\cdot p')
\nonumber\\
&&
-\left(m_n^{2}+p\cdot p'\right)q^2
\Big]\delta^{ik}_{\perp}.
\eea
%-------------------------
Here $\delta^{ik}_\perp=\delta^{ik}-q^iq^k/q^2$.
The expressions~\eqref{eq:H_00}--\eqref{eq:H_ik_perp} simplify in the
nonrelativistic limit $m_n\gg |\vecp|,|\vecp'|\simeq |\vecq|$ for
neutrons. Approximating
$\ep_p\simeq m_n^*+\vecp^{\,2}/2m_n$, and keeping only the leading
terms, we obtain
%-----------------------
\bea\label{eq:H_00_approx}
H^{00}
&\simeq&
2(\vecp+\vecp')^2\vecq^{\,2}
-8(\vecq\cdot\vecp)(\vecq\cdot\vecp'),
\\
\label{eq:H_0i_perp_approx}
H^{0i}_\perp
&\simeq&
4m_n\vecq^{\,2}
\left(p^{i}_\perp+p'^{i}_\perp\right),
\nonumber\\
H^{ik}_\perp
&\simeq&
8m_n^{2}\vecq^{\,2}\delta^{ik}_{\perp}.
\eea
%-------------------------

%-----------------------------------------------------------
\begin{figure}[t]
\centering
\includegraphics[width=\columnwidth,keepaspectratio]{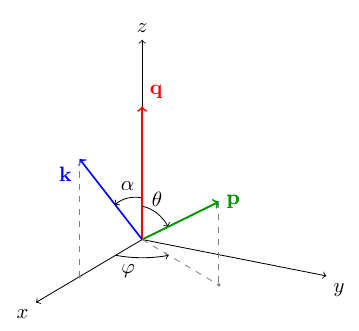}
\caption{Relative orientation of the three vectors $\vecq$, $\veck$,
and $\vecp$, with angles defined in Eq.~\eqref{eq:angles}.}
\label{fig:angles}
\end{figure}
% -----------------------------------------------------------

Finally, taking the ultrarelativistic limit $k,k'\gg m_e$, so that
$\epsilon_k\simeq k$ and $\epsilon_{k'}\simeq k'$, the matrix element
is obtained from Eqs.~\eqref{eq:L_00}--\eqref{eq:L_ik} together with
Eqs.~\eqref{eq:H_00_approx}--\eqref{eq:H_0i_perp_approx}. This gives
%-------------------------
\bea\label{eq:L_00H_00}
L^{00}H^{00}
&=&
8\left(kk' + \veck \cdot\veck'\right)
\nonumber\\
&&
\times
\left[
(\vecp+\vecp')^2\vecq^{\,2}
-4(\vecq\cdot\vecp)(\vecq\cdot\vecp')
\right],
\\
\label{eq:L_0iH_0i}
L^{0i}_\perp H^{0i}_\perp
&=&
32\vecq^{\,2}m_n(k+k')
\left(\veck_\perp\cdot \vecp_\perp\right),
\\
\label{eq:L_ikH_ik}
L^{ik}_\perp H^{ik}_\perp
&=&
64\vecq^{\,2}m_n^{2}
\left(
\veck_\perp^2
-\veck\cdot\veck'
+kk'
\right),
\eea
%-------------------------
where we used
$k'^i_\perp=k^i_\perp$, $p'^i_\perp=p^i_\perp$, and
$\delta^{ik}_\perp\delta^{ik}_\perp=2$.

To proceed, we choose a Cartesian coordinate system with $\vecq$ fixed
along the $z$-axis, as shown in Fig.~\ref{fig:angles}. The relevant
scalar products are
%-------------------------
\bea\label{eq:angles}
\veck\cdot\vecq
&=&
kq\cos\alpha,
\qquad
\vecp\cdot\vecq
=
pq\cos\theta,
\nonumber\\
\veck_\perp\cdot\vecp_\perp
&=&
k_\perp p_\perp\cos\varphi,
%\nonumber\\
\quad k_\perp
=
k\sin\alpha,
\qquad
p_\perp
=
p\sin\theta,
\nonumber\\
\veck\cdot\veck'
&=&
k^2-kq\cos\alpha,
%\nonumber\\
\quad \vecq\cdot\vecp'
=
q^2+pq\cos\theta,
\nonumber\\
(\vecp+\vecp')^2
&=&
(2\vecp+\vecq)^2
=
4p^2+q^2+4pq\cos\theta,
\eea
%------------------------
where $\veck'=\veck-\vecq$, $\vecp'=\vecp+\vecq$, and $k'=k-\omega$.
Using these relations in Eqs.~\eqref{eq:L_00H_00}--\eqref{eq:L_ikH_ik}
yields
%-------------------------
\bea\label{eq:angL_00H_00}
L^{00}H^{00}
&=&
8q^2k
\left(2k-\omega-q\cos\alpha\right)
\left(4p^2\sin^2\theta+q^2\right),\quad
\\
\label{eq:angL_0iH_0i}
L^{0i}_\perp H^{0i}_\perp
&=&
32m_nkpq^2
(2k-\omega)
\sin\alpha\sin\theta\cos\varphi,
\\
\label{eq:angL_ikH_ik}
L^{ik}_\perp H^{ik}_\perp
&=&
64q^2m_n^{2}k
\left(
k\sin^2\alpha+q\cos\alpha-\omega
\right),
\eea
%-------------------------
where $q^2\equiv |\vecq|^2$. Substituting
Eqs.~\eqref{eq:angL_00H_00}--\eqref{eq:angL_ikH_ik} and
Eqs.~\eqref{eq:matrix_en3}--\eqref{eq:matrix_en5} into
Eq.~\eqref{eq:matrix_en1}, we obtain
%---------------------------------------
\bea\label{eq:matrix_en7}
\overline{|\mathcal{M}^{en}_{12\to 34}|^2}
&=&
\frac{e^4\kappa_n^2}
{32m_n^{4}(k-\omega)}
\Bigg[
\frac{A(\alpha,\theta)}
{|q^2+\Pi_L|^2}
\nonumber\\
&-&
{\rm Re}
\frac{B(\alpha,\theta)\cos\varphi}
{(q^2+\Pi_L)
(q^2-\omega^2+\Pi_T^{*})}
\nonumber\\
&&
+
\frac{C(\alpha)}
{|q^2-\omega^2+\Pi_T|^2}
\Bigg],
\eea
%-------------------------
where
%-------------------------
\bea\label{eq:A_func}
A(\alpha,\theta)
&=&
q^2
\left(2k-\omega-q\cos\alpha\right)
\left(4p^2\sin^2\theta+q^2\right),\quad
\\
\label{eq:B_func}
B(\alpha,\theta)
&=&
8m_npq^2(2k-\omega)
\sin\alpha\sin\theta,
\\
\label{eq:C_func}
C(\alpha)
&=&
8q^2m_n^{2}
\left(k\sin^2\alpha+q\cos\alpha-\omega\right).
\eea
%-------------------------

\section{Electron-neutron scattering rate}
\label{app:relax_en}

To evaluate the electron-neutron scattering rate, we start from
Eq.~\eqref{eq:t_relax_en}. Introducing a dummy integration over the
transferred energy
$\omega=\ep_1-\ep_3=\ep_4-\ep_2$, we find
%-----------------------------------
\bea\label{eq:t_relax_en1}
\tau_{en}^{-1}(\ep)
&=&
(2\pi)^{-5}
\int_{-\infty}^{\infty} d\omega
\int d\vecq
\int d\vecp'\,
\overline{|{\cal M}^{en}_{12\to 34}|^2}
\nonumber\\
&&
\times
\frac{\bm q\cdot \bm p}{p^2}
\delta(\ep-\ep_3-\omega)
\delta(\ep'-\ep_4+\omega)
\nonumber\\
&&
\times
\frac{1-f^0_3}{1-f_1^{0}}
f'_2(1-f'_4).
\eea
%-----------------------------------
The energies $\ep_3$ and $\ep_4$ should be evaluated at
$\vecp_3=\vecp-\vecq$ and $\vecp_4=\vecp'+\vecq$, respectively, where
we have redefined
$\bm p_1=\bm p$, $\bm p_2=\bm p'$, $\ep_1=\ep$, and $\ep_2=\ep'$.
Substituting Eq.~\eqref{eq:matrix_en7} into Eq.~\eqref{eq:t_relax_en1}
gives
%-----------------------------------------------
\bea\label{eq:t_relax_en2}
\tau^{-1}_{en}
&=&
\frac{m_n}{(2\pi)^5p}
\int_{-\infty}^{\infty} d\omega
\int d\vecq\,
\cos\alpha\,
\delta(\epsilon-\epsilon_3-\omega)
\nonumber\\
&&
\times
\frac{1-f^0_3}{1-f^0_1}
\int_{0}^\infty dp'\,p'\,
f'_2(1-f'_4)\,
I_\Omega ,
\eea
%-----------------------------------------------
where the angular integral is
%-----------------------------------------------
\bea\label{eq:angle_integral_new}
I_{\Omega}
&=&
\frac{(2\pi)e^4\kappa_n^2}
{32m_n^4(p-\omega)}\nonumber\\
&\times&
\Bigg[
\frac{\tilde{A}(\alpha)}
{|q^2+\Pi_L|^2}
+
\frac{C(\alpha)}
{|q^2-\omega^2+\Pi_T|^2}
\Bigg]
\theta(1-|x_0|),
\eea
%-----------------------------------------------
with
%-----------------------------------------------
\bea\label{eq: A_tilde}
\tilde{A}(\alpha)
&=&
q^2(2p-\omega-q\cos\alpha)
\left[
q^2+4p'^2-4(p'x_0)^2
\right].\nonumber\\
\eea
% -----------------------------------------------
The term proportional to $\cos\varphi$ vanishes upon angular
integration.

The $\theta$-function in Eq.~\eqref{eq:angle_integral_new} defines the
minimum value of $p'$ as $|p_0|$, with
$p_0=(2\omega m_n-q^2)/(2q)$, which also implies $x_0p'=p_0$.
Substituting Eq.~\eqref{eq:angle_integral_new} into
Eq.~\eqref{eq:t_relax_en2}, we obtain
%--------------------------------------------
\bea\label{eq:t_relax_en3}
\tau^{-1}_{en}
&=&
\frac{e^4\kappa_n^2}
{32(2\pi)^3m_n^2p^2}
\int_{-\infty}^{\epsilon}d\omega\,
\frac{\epsilon-\omega}{p-\omega}
\nonumber\\
&&
\times
\frac{1-f^0(\epsilon-\omega)}
{1-f^0(\epsilon)}
\int_0^\infty qdq\,
y_0\theta(1-|y_0|)
\nonumber\\
&&
\times
\int_{\ep_{\rm min}}^\infty d\ep'\,
f'(\epsilon')
\left[1-f'(\epsilon'+\omega)\right]
\nonumber\\
&&
\times
\Bigg[
\frac{\bar{A}}{|q^2+\Pi_L|^2}
+
\frac{\bar{C}}{|q^2-\omega^2+\Pi_T|^2}
\Bigg],
\eea
%-----------------------------------------------
where $p'dp'=m_nd\ep'$ and $\ep_{\rm min}=p_0^2/2m_n$. The functions
$\bar{A}$ and $\bar{C}$ are
%-----------------------------------------------
\bea\label{eq:A_bar}
\bar{A}
&=&
q^2
\left(2p-\omega-qy_0\right)
\left(q^2+8m_n\ep'-4p_0^2\right)
\nonumber\\
&\simeq&
\frac{2m_n}{\ep}
\left[(2\ep-\omega)^2-q^2\right]
\left(
2q^2\ep'
-\omega^2m_n
+\omega q^2
\right),
\\
\label{eq:C_bar}
\bar{C}
&=&
8q^2m_n^2
\left[
p(1-y_0^2)+qy_0-\omega
\right]
\nonumber\\
&\simeq&
\frac{2m_n^2}{\ep}
\Big[
q^4
+4q^2\ep(\ep-\omega)
-\omega^2(2\ep-\omega)^2
\Big],
\eea
%-----------------------------------------------
where $p_0$ and $y_0$ have been substituted, and $p\simeq\ep$ was used in the second step.

%-------------------------------------------------------
\begin{figure*}[t]
\centering
\includegraphics[width=0.45\textwidth,keepaspectratio]{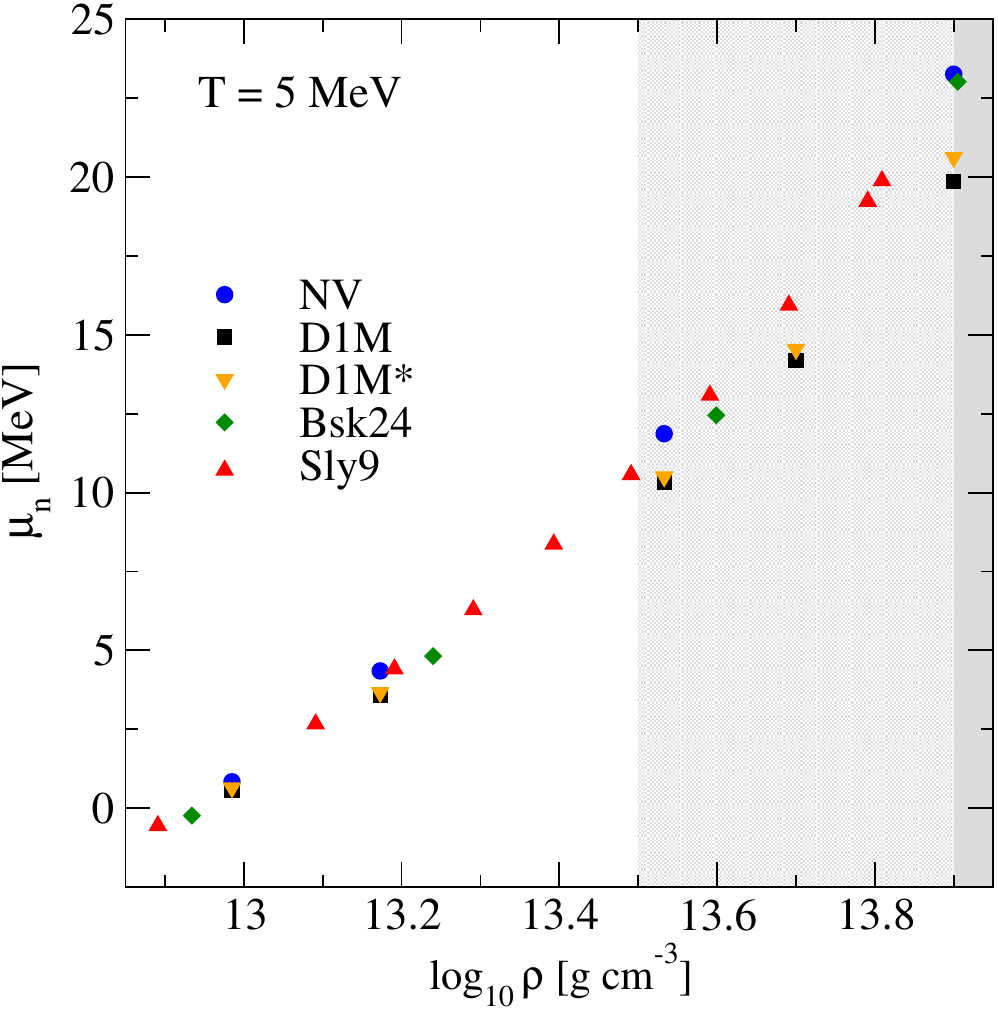}
\hspace{0.5cm}
\includegraphics[width=0.45\textwidth,keepaspectratio]{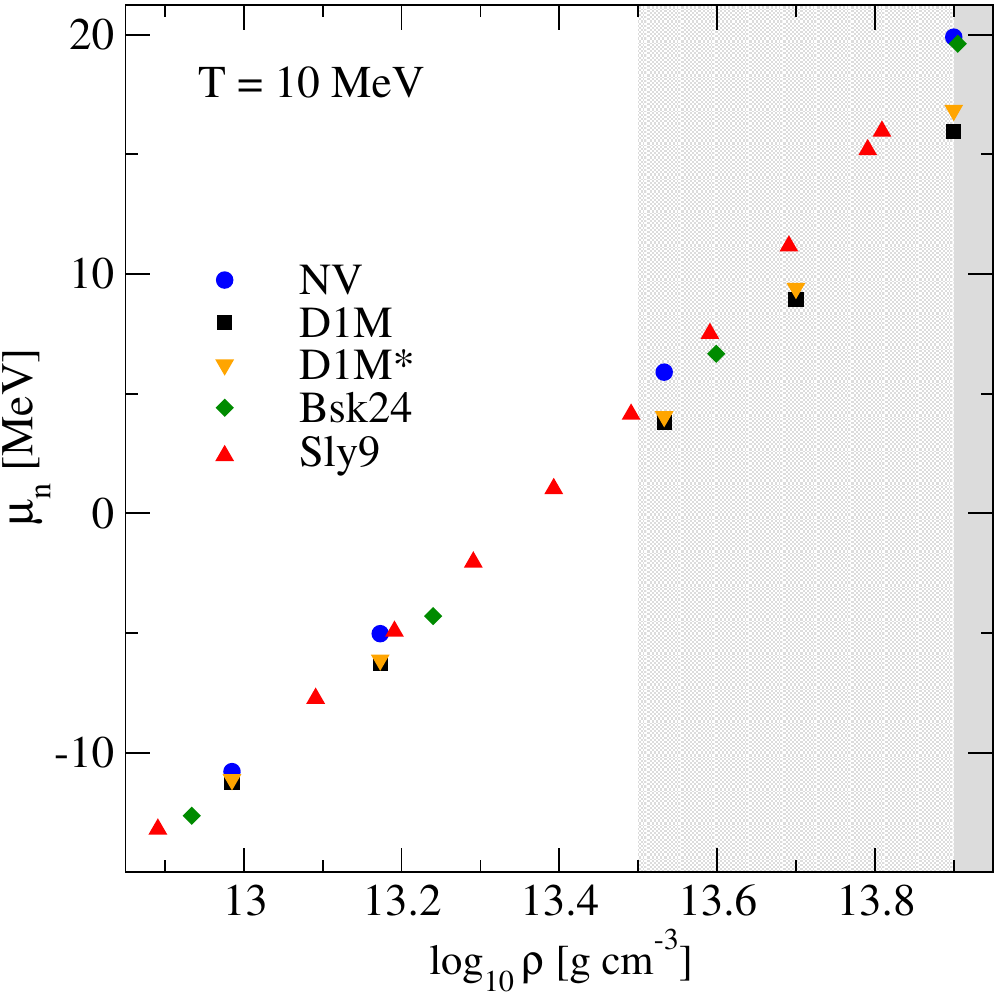}
\caption{The chemical potential $\mu_n$ of a free neutron gas evaluated
at temperature $T=5$ MeV left and $T=10$ MeV right as a function of
density for five compositions. The shaded regions are the same as in
Fig.~\ref{fig:radii}.}
\label{fig:mu_n_dens}
\end{figure*}
%-------------------------------------------------------

The $\theta$-function in Eq.~\eqref{eq:t_relax_en3} restricts the
$q$-integration to $q_-\le q\le q_+$, where
$q_{\pm}=|\sqrt{(\ep-\omega)^2-m^2}\pm\sqrt{\ep^2-m^2}|$.
To keep $q$ real, one must also impose $\omega\le \ep-m_e$.
Implementing these limits, we obtain
%----------------------------------------------
\bea\label{eq:t_relax_en5}
\tau^{-1}_{en}(\ep)
&=&
\frac{\alpha^2\kappa_n^2}
{16\pi m_n\ep^4}
\int_{-\infty}^{\epsilon-m}d\omega\,
\frac{1-f^0(\epsilon-\omega)}
{1-f^0(\epsilon)}
\nonumber\\
&&
\times
\int_{q_-}^{q_+}dq\,
\left(q^2-\omega^2+2\ep\omega\right)
\nonumber\\
&&
\times
\int_{\ep_{\rm min}}^\infty d\ep'\,
f'(\epsilon')
\left[1-f'(\epsilon'+\omega)\right]
\nonumber\\
&&
\times
\Bigg[
\frac{N_L}{|q^2+\Pi_L|^2}
+
\frac{N_T}{|q^2-\omega^2+\Pi_T|^2}
\Bigg],
\eea
%-----------------------------------------------
where $e^2=4\pi\alpha$ and
%-----------------------------------------------
\bea\label{eq:NL}
N_L
&=&
\left[(2\ep-\omega)^2-q^2\right]
\left[
q^2(2\ep'+\omega)-\omega^2m_n
\right],
\\
\label{eq:NT}
N_T
&=&
m_n
\left[
q^4
+4q^2\ep(\ep-\omega)
-\omega^2(2\ep-\omega)^2
\right].
\eea
%-----------------------------------------------

To evaluate the relaxation time~\eqref{eq:t_relax_en5}, we first
determine the chemical potential from the local density of free
neutrons $n_n$, continuing to assume $m_n^*\simeq m_n$, \ie, a free
Fermi gas. The resulting density dependence of the neutron chemical
potential at $T=5$~MeV and $T=10$~MeV is shown in
Fig.~\ref{fig:mu_n_dens}. Neutrons remain nondegenerate at
$T=10$~MeV, or only semidegenerate at $T=5$~MeV, up to densities of
order $\rho\simeq 10^{13.5}$~g\,cm$^{-3}$, becoming fully degenerate
only near the bottom of the inner crust. The case $T=1$~MeV is similar
to the case $T=5$~MeV and is not shown. At low densities, all models
yield nearly identical values of $\mu_n$. However, above
$\rho\simeq 10^{13.5}$~g\,cm$^{-3}$, small differences begin to appear,
growing to about 10~MeV at densities around
$10^{14}$~g\,cm$^{-3}$. These deviations arise because, for
compositions with smaller nuclei, the number of free neutrons in the
Wigner--Seitz cell increases more rapidly than the free-neutron volume
itself, leading to larger $n_n$ and hence larger $\mu_n$.

%------------------------------------------------------
\begin{figure*}[t]
\centering
\includegraphics[width=0.45\textwidth,keepaspectratio]
{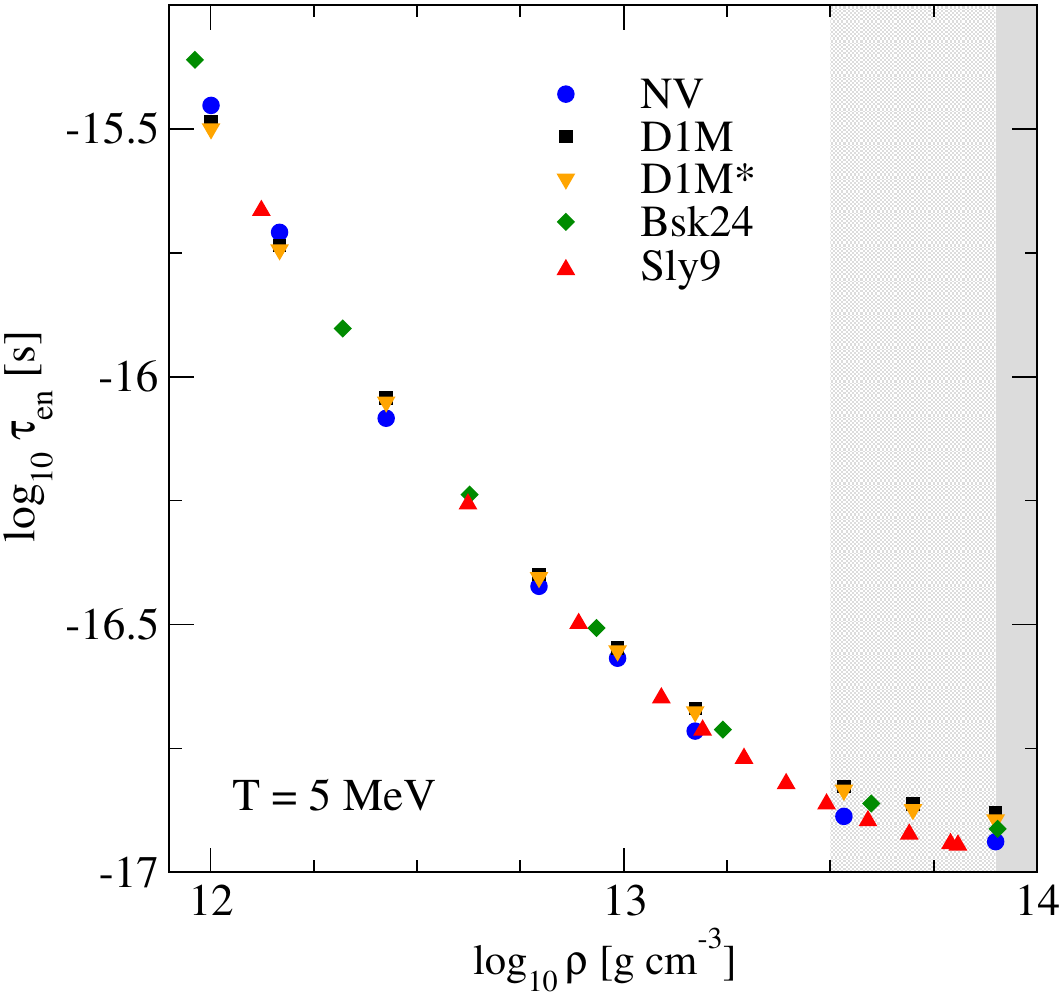}
\hspace{0.5cm}
\includegraphics[width=0.45\textwidth,keepaspectratio]
{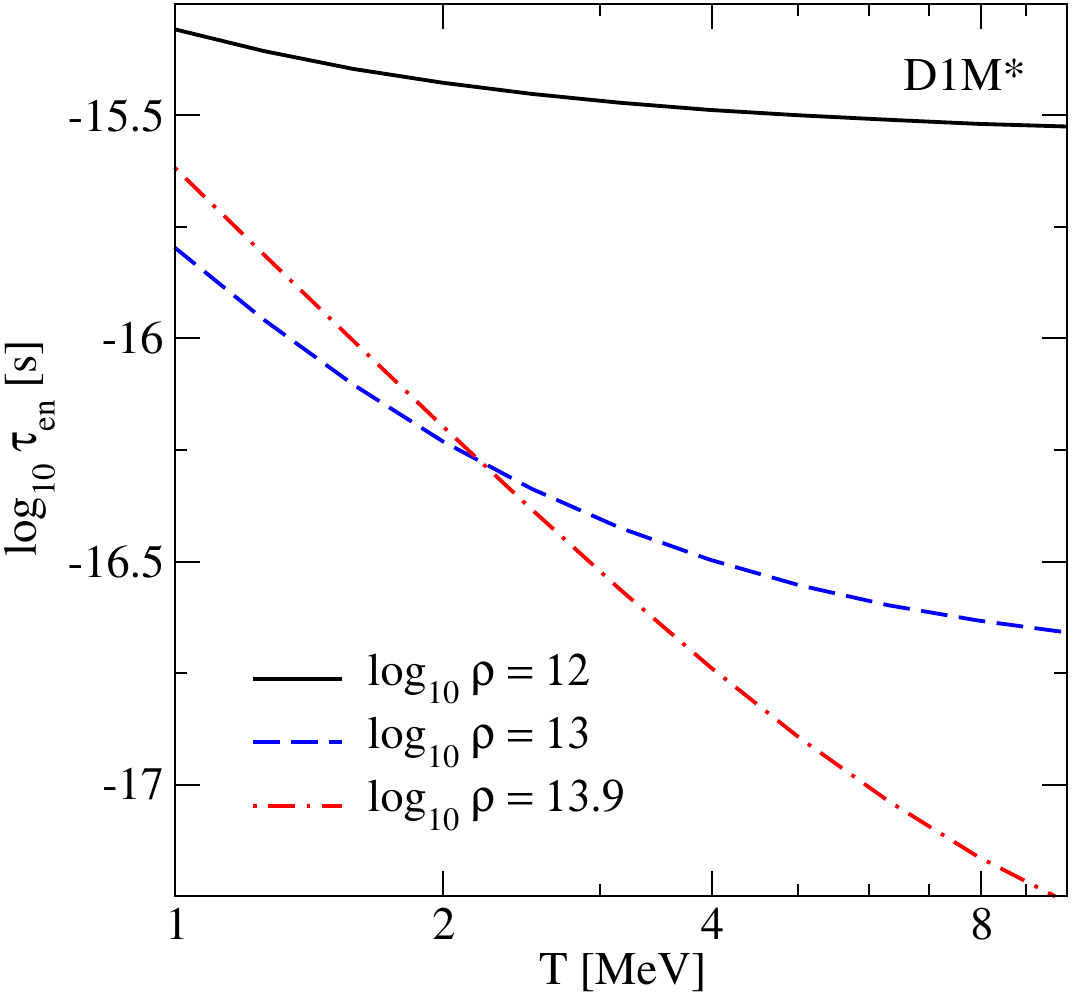}
\caption{Left panel: Relaxation time $\tau_{en}$ as a function of
  density for various compositions at temperature $T=5$~MeV. Right
  panel: Temperature dependence of $\tau_{en}$ for various densities
  for model D1M*. The shaded regions are the same as in
  Fig.~\ref{fig:radii}.}
\label{fig:tau_en}
\end{figure*}
%--------------------------------------------------------

Figure~\ref{fig:tau_en} left shows the density dependence of the
electron--neutron scattering relaxation time at $T=5$~MeV. The
relaxation time decreases with increasing density, in contrast to the
electron--ion case. This trend reflects the growing free-neutron
fraction at higher densities. Unlike electron--ion scattering, where
the nuclear form factor becomes important at high densities,
electron--neutron scattering is unaffected by finite-size effects of
the neutron. The weak dependence on nuclear composition stems from the
small variation in $\mu_n$ among the models: compositions with larger
neutron chemical potentials correspond to higher neutron densities and
therefore to shorter $\tau_{en}$, as expected.

It is also worth mentioning that the main contribution to
electron-neutron scattering comes from the transverse part of the matrix
element, in contrast to electron-ion scattering. This follows directly
from Eqs.~\eqref{eq:NL} and \eqref{eq:NT}, which imply
$N_T\gg N_L$. This is expected because electron-neutron scattering is
mediated by a magnetic, \ie, transverse, interaction, whereas
electron-ion scattering is dominated by Coulomb electrostatic forces.
Numerically, we find that the longitudinal part contributes only a few
percent to the total electron-neutron scattering rate.

The temperature dependence of the electron-neutron scattering
relaxation time is shown in the right panel of Fig.~\ref{fig:tau_en}
for three different densities. The results are shown for the
composition D1M*, but they are very similar for the other models. The
relaxation time decreases as the temperature increases, with a stronger
decrease at higher densities. This behavior mainly reflects the fact
that increasing temperature raises the typical transferred plasmon
energy $\omega$, which increases the scattering rate and therefore
shortens the relaxation time. In addition, at higher densities, the
larger number of scattering centers leads to a more pronounced decrease
in the relaxation time.

Comparing Fig.~\ref{fig:tau_dens} with the left panel of
Fig.~\ref{fig:tau_en}, we find that $\tau_{en}\gg\tau_{ei}$ throughout
the inner crust, with the smallest ratio
$\tau_{en}/\tau_{ei}\sim 10^2$ occurring in the densest layers.
Consequently, electron-neutron scattering has a negligible impact on
electron transport across the crust. An analogous conclusion was
reached in Ref.~\cite{Flowers1976} for cold neutron-star matter in the
solid phase.

\section{Low-temperature limit of transport coefficients}
\label{app:low-T}

To derive the low-temperature counterparts of the transport
coefficients, we use the standard expansion formula for $T\ll T_F$,
%-----------------------------------
\bea\label{eq:expansion}
\int_m^\infty d\ep\,
\frac{\partial f^0}{\partial\ep}
F(\ep)
&=&
-F(\mu)
-\frac{\pi^2T^2}{6}F''(\mu)
\nonumber\\
&&
+{\cal O}(T^4),
\eea
%-----------------------------------
where $F(\ep)$ is a well-behaved function around $\ep=\mu$, and primes
denote derivatives with respect to $\ep$. For the tensor in
Eq.~\eqref{eq:cond_tensors} with $n=0$, it is sufficient to keep the
first term with $\mu\simeq\ep_F$. For the tensors with $n=1,2$, the
first term vanishes, and the leading term is of order $T^2$. Thus,
%-----------------------------------
\bea\label{eq:L_0_deg}
{\cal L}^{0}_{l}
=
\frac{1}{3\pi^2}
{\cal F}_l(\ep_F),
\quad 
{\cal L}^{1}_{l}
=
\frac{T}{9}{\cal F}'_l(\ep_F),
\quad 
{\cal L}^{2}_{l}
=
\frac{1}{9}{\cal F}_l(\ep_F).\nonumber\\
\eea
% -----------------------------------
To evaluate ${\cal F}'_l(\ep)$, we use
$p\simeq\ep$, $\tau\propto\ep^2$~\cite{Harutyunyan2016}, and
$\omega_c\propto\ep^{-1}$. Hence
$\tau'=2\tau/\ep$ and
$(\omega_c\tau)'=\omega_c\tau/\ep$. We find
%-----------------------------------
\bea\label{eq:F_deriv}
{\cal F}'(\ep)
&=&
\frac{d}{d\ep}
\left(
\frac{p^3}{\ep}\tau
\right)
=
4p\tau,
\\
\label{eq:F0_deriv}
{\cal F}'_0(\ep)
&=&
\frac{d}{d\ep}
\left[
\frac{p^3}{\ep}
\frac{\tau}{1+(\omega_c\tau)^2}
\right]
\nonumber\\
&=&
2p\tau
\frac{2+(\omega_c\tau)^2}
{\left[1+(\omega_c\tau)^2\right]^2},
\\
\label{eq:F1_deriv}
{\cal F}'_1(\ep)
&=&
\frac{d}{d\ep}
\left[
\frac{p^3}{\ep}
\frac{\tau(\omega_c\tau)}
{1+(\omega_c\tau)^2}
\right]
\nonumber\\
&=&
p\omega_c\tau^2
\frac{5+3(\omega_c\tau)^2}
{\left[1+(\omega_c\tau)^2\right]^2}.
\eea
%-----------------------------------
Equations~\eqref{eq:cond_tensors} and \eqref{eq:L_0_deg} lead to the
low-temperature limiting forms of the components of the electrical
conductivity tensor,
%-----------------------------------
\bea\label{eq:sigma_l_deg}
\sigma_l
&=&
\frac{e^2}{3\pi^2}{\cal F}_l(\ep_F)
=
\frac{e^2}{3\pi^2}
\frac{p_F^3}{\ep_F}\tau_F
\frac{(\omega_{cF}\tau_F)^l}
{1+(\omega_{cF}\tau_F)^2}
\nonumber\\
&=&
\frac{n_ee^2\tau_F}{\ep_F}
\frac{(\omega_{cF}\tau_F)^l}
{1+(\omega_{cF}\tau_F)^2},
\\
\label{eq:sigmas_deg}
\sigma
&=&
\frac{n_ee^2\tau_F}{\ep_F},\quad 
\sigma_0
=
\frac{\sigma}
{1+(\omega_{cF}\tau_F)^2},
\nonumber\\
\sigma_1
&=&
\frac{\omega_{cF}\tau_F}
{1+(\omega_{cF}\tau_F)^2}\sigma,
\eea
%-----------------------------------
where $p_F^3=3\pi^2n_e$ and
$\omega_{cF}=\omega_c(\ep_F)$, $\tau_F=\tau(\ep_F)$. The first
expression in Eq.~\eqref{eq:sigmas_deg} is the Drude formula. For
$\tilde{\kappa}_l$ and $\alpha_l$, Eq.~\eqref{eq:cond_tensors} gives
%-----------------------------------
\bea\label{eq:kappa_tilde_alpha_l_deg}
\tilde{\kappa}_{l}
&=&
\frac{T}{9}{\cal F}_l(\ep_F)
=
\frac{\pi^2}{3e^2}T\sigma_l,
\quad 
\alpha_{l}
=
-\frac{eT}{9}{\cal F}'_l(\ep_F).
\eea
%-----------------------------------
Using Eqs.~\eqref{eq:F_deriv}--\eqref{eq:F1_deriv}, we obtain
%-----------------------------------
\bea\label{eq:kappa_tilde_deg}
\tilde\kappa
&=&
\frac{\pi^2 n_e\tau_F}{3\ep_F}T,
\quad 
\tilde\kappa_0
=
\frac{\tilde\kappa}
{1+(\omega_{cF}\tau_F)^2},
\nonumber\\
\tilde\kappa_1
&=&
\frac{\omega_{cF}\tau_F}
{1+(\omega_{cF}\tau_F)^2}\tilde\kappa .\nonumber\\
\eea
% -----------------------------------
For $\alpha_l$, one finds
%-----------------------------------
\bea\label{eq:alpha_deg}
\alpha
&=&
-\frac{eT}{9}(3+v_F^2)p_F\tau_F
=
-\frac{4eT}{9}p_F\tau_F,
\\
\label{eq:alpha0_deg}
\alpha_0
&=&
-\frac{eT}{9}p_F
\frac{\tau_F}
{1+(\omega_{cF}\tau_F)^2}
\nonumber\\
&&
\times
\left[
2+
\frac{2}
{1+(\omega_{cF}\tau_F)^2}
\right],
\\
\label{eq:alpha1_deg}
\alpha_1
&=&
-\frac{eT}{9}p_F
\frac{\tau_F(\omega_{cF}\tau_F)}
{1+(\omega_{cF}\tau_F)^2}
\nonumber\\
&&
\times
\left[
3+
\frac{2}
{1+(\omega_{cF}\tau_F)^2}
\right].
\eea
%-----------------------------------
From Eqs.~\eqref{eq:sigmas_deg}, the conductivity components satisfy
%-----------------------------------
\bea\label{eq:sigma_rel}
\sigma_0^2+\sigma_1^2
&=&
\left[
\frac{1}{[1+(\omega_{cF}\tau_F)^2]^2}
+
\frac{(\omega_{cF}\tau_F)^2}
{[1+(\omega_{cF}\tau_F)^2]^2}
\right]\sigma^2
\nonumber\\
&=&
\frac{\sigma^2}
{1+(\omega_{cF}\tau_F)^2}
=
\sigma\sigma_0 .
\eea
%-----------------------------------
For the components of the thermopower $Q$, using
Eqs.~\eqref{eq:Q_comp} and
\eqref{eq:alpha_deg}--\eqref{eq:alpha1_deg}, we find
%-----------------------------------
\bea\label{eq:Q_deg1}
Q
&=&
\frac{4\pi^2T}{3e\ep_F},
\\
\label{eq:Q_deg2}
Q_0
&=&
\frac{\pi^2T}{3e\ep_F}
\left[
3+
\frac{1}
{1+(\omega_{cF}\tau_F)^2}
\right],
\\
\label{eq:Q_deg3}
Q_1
&=&
\frac{\pi^2T}{3e\ep_F}
\frac{\omega_{cF}\tau_F}
{1+(\omega_{cF}\tau_F)^2}.
\eea
%-----------------------------------
For small Hall parameter, $\omega_{cF}\tau_F\ll1$, one obtains
%-----------------------------------
\bea\label{eq:Q_isotrop1}
Q_0
&\simeq&
Q,
\\
\label{eq:Q_isotrop2}
Q_1
&\simeq&
\frac{\pi^2T}{3e\ep_F}
(\omega_{cF}\tau_F)
\nonumber\\
&=&
\frac{\pi^2\tau_F}{3\ep_F^2}TB
=
\frac{1}{4}(\omega_{cF}\tau_F)Q
\ll Q,
\eea
%-----------------------------------
whereas in the opposite limit, $\omega_{cF}\tau_F\gg1$,
%-----------------------------------
\bea\label{eq:Q_anisotrop1}
Q_0
&\simeq&
\frac{p_F\ep_F}{3n_ee}T
=
\frac{3Q}{4},
\\
\label{eq:Q_anisotrop2}
Q_1
&\simeq&
\frac{\pi^2T}{3e\ep_F}
\frac{1}{\omega_{cF}\tau_F}
\nonumber\\
&=&
-\frac{\pi^2}{3e^2\tau_F}
\frac{T}{B}
=
\frac{1}{4}
\frac{Q}{\omega_{cF}\tau_F}
\ll Q.
\eea
%-----------------------------------
Using Eqs.~\eqref{eq:kappa_scalar}--\eqref{eq:kappa_1} and
\eqref{eq:kappa_tilde_deg}--\eqref{eq:Q_deg3}, we find that in the
low-$T$ limit $\hat{\kappa}\simeq\hat{\tilde{\kappa}}$. From
Eq.~\eqref{eq:kappa_tilde_alpha_l_deg}, one recovers the
Wiedemann--Franz law for low-temperature degenerate matter,
$(3e^2\kappa)/(\pi^2\sigma T)\to1$. For
$\omega_{cF}\tau_F\ll1$, one finds
%-----------------------------------
\bea\label{eq:kappa_isotrop}
\kappa_0
&\simeq&
\kappa,
\quad 
\kappa_1
\simeq
(\omega_{cF}\tau_F)\kappa
=
\frac{\pi^2 n_ee\tau_F^2}{3\ep_F^2}TB
\nonumber\\
&\simeq&
\frac{3eB}{\pi^2n_eT}\kappa^2,
\eea
%-----------------------------------
whereas for $\omega_{cF}\tau_F\gg1$,
%-----------------------------------
\bea\label{eq:kappa_anisotrop}
\kappa_0
&\simeq&
\frac{\kappa}{(\omega_{cF}\tau_F)^2}
=
\frac{\pi^2n_e\ep_F}{3e^2\tau_F}
\frac{T}{B^2}
=
\left(
\frac{\pi^2n_eT}{3eB}
\right)^2
\kappa^{-1},\nonumber\\
\\
\kappa_1
&\simeq&
\frac{\kappa}{\omega_{cF}\tau_F}
=
\frac{\pi^2n_eT}{3eB}.
\eea
% -----------------------------------

% The sn-mathphys-num class option supplies the Springer Nature
% numerical mathematics/physics reference style.
%\bibliographystyle{JHEP}
%\bibliography{CrustTrans}

\providecommand{\href}[2]{#2}\begingroup\raggedright\endgroup

\end{document}